\pdfoutput=1
\documentclass{aastex62}

\begin{document}

\title { A Model For Dark Matter Haloes.}
\title {II. A  GR Source Term $T^{\mu\nu}$ For Dark Matter.}

\author{H. L.  Helfer}
\email{lary@astro.pas.rochester.edu}
\affiliation{ Department of Physics and Astronomy,  The University of Rochester,}
\affiliation{Rochester, NY 14627}
\date{\today}

\begin{abstract}
         
         Two forms are suggested for the source term $T^{\mu \nu}$ associated with an aggregate of dark matter (with the properties described in Paper I). Both  have   large pressure-like components which dominate the density terms. Using one form a simple model of the spiral galaxy  halos is developed which can match the observed `flat'  outer rotation curves of some galaxies,including  the Milky Way.  It can also represent the ascending outer rotational curves of small spiral galaxies such as M33.  See Figs. 1, 2. 
         
         The analysis of the Milky Way rotation curve  gives $p_a/c^2 \sim 6 \times 10^{-27} \ {\rm g \ cm}^{-3}$  at $R \sim 60-80$ kpc  near the outer edge of the halo.  The surrounding  dark matter (DM) cloud has $p/c^2 \sim  0.1 p_a/c^2$.  The dark matter cannot come closer to the center than $R \sim 4$ kpc.  It reaches a maximum $p_b/c^2 \sim  10^2 p_a/c^2$ at $R_h \simeq 8$ kpc and then falls rapidly $p/c^2 \propto 1/r^2$.  It is hoped that  by analyzing other galactic rotation curves these models can be used to infer properties of the intergalactic DM matter. 
         
           The other form for $T^{\mu \nu}$ for DM is useful in cosmology.  Then a (non-constant) cosmological `constant' term $\Lambda$ needs to be added to Einstein's equation in order to allow use of  the standard $p\ - \rho$ relation for DM.   We suggest that at least some of the `dark energy'  present may have been misinterpreted and not be real, resulting from  adopting an improper   form of   $T^{\mu \nu}$  for DM.
     

\end{abstract}


\section{Introduction}
\label{Sec:1}
\subsection{Preliminary Concepts.}

      In this  paper, dark matter (DM) refers to the  unknown  material responsible for the extended gravitational fields needed in constructing models for gravitational lensing,  for  modeling the motions of ordinary matter (OM) in clusters of galaxies and in the halos of many spiral galaxies, including the Milky Way.  
      
      For these environments,  a model for DM  was proposed  which assumed it is comprised of particles that  follow time-like world-lines similar to those of  non-relativistic  OM particles.  However,  the DM particles have a different kinematics than OM particles have, being influenced by their internal motions, so that their total momenta is space-like.  See Paper I (\cite {H1}). 
      
      [This  model of DM  may also represent the `cold' DM encountered in the $\Lambda CDM$ cosmology.  We do not  (and do not need to) speculate upon the particle physics nature of the constituents of  DM.    We also noted that speculation about  the classification of internal motions  within fundamental particles is appropriate; a recent calculationattributes $\sim 68\%$ of the mass of the proton to internal motions. ]
       
              As emphasized in \cite{H1}, an important observational property of dark matter (DM) is that it does not appear  in substantial amounts in ordinary stars.  As a consequence, a suggestion  was made that DM objects were characterized by having their total momenta  being space-like, reflecting the presence  of two different types of internal momentum flows: (1) a time-like component $\hat U$ which indirectly specifies its motion through space-time; and (2), a space-like component $\hat S$ which is transported with the object.  The particle actually moves along a complicated path whose tangent is the vector $U$. The path is a tight helix whose axis  is the particle's world-line, described by a tangent vector  $\langle \hat U \rangle$; it satisfies $\langle \hat U^\nu \rangle \langle \hat U^\mu \rangle_{;\nu} =0$, the usual geodesic equation. It was shown that part of $\hat U$ appears as an internal spin-like motion which can couple to the object's orbital angular momentum; this coupling creates a centrifugal barrier which prevents the DM object  from penetrating  into  deep gravitational potential wells such as those around proto-stars.    This explains its absence in stars.

            One component of the internal field $\hat S$  is needed to provide strict local conservation of momentum for balancing the helical component of $\hat U$.  The other part\footnote{It is referred to in text   and in Paper I as $\hat S^z$.} of  $\hat S$  would be carried along and contributes to the  local DM energy-momentum tensor $T^{\mu \nu}$, a source term for Einstein's equation of GR. It is relatively large, because the total momentum is taken to be space-like.   This paper deals with the form of this contribution when dealing with aggregates of these DM particles. Like rest-mass, this DM aggregate momentum flux contributes significantly to the gravitational field.
            
               A very simple DM model of the halo region of a galaxy  is constructed by studying  the effects of angular momentum conservation on determining the distribution of the halo DM.  The model's utility  is tested by requiring it to explain the outer part of galactic rotation curves, commonly attributed to DM halos.   Successful models for the large spiral galaxy, the Milky Way, and the small spiral  M33, are presented. [See Figs. 1 \& 2].

               Appendix A summarizes the observations, primarily for the large spirals, exhibiting `flat' rotation curves. The notation used  in this paper is that of Paper I.\footnote{ One has $c=1$ normally and  the local Lorentz metric $g_{\mu\nu}$has the signature $(+1,-1,-1,-1)$. Also the time-like momentum vector is written as $\hat U =m_0 U$ and similarly for the space-like momentum vector $\hat S =m_1S$. }

\subsection{Outline.}  

  \subsubsection{ The DM Source Term $T^{\mu\nu}$}  
       
       In many GR problems the energy-momentum tensor $T^{\mu\nu}$, used as a source term, is taken to be that of a classical fluid.  First, we start with examining some problems of using a kinetic theory representation of matter to form a fluid energy-momentum tensor  primarily in order to  understand what the  symbol  $p$ means; there is a distinction between thermodynamic pressure terms and other forms of momentum flux.  In general in this paper $p$ is not a thermodynamic pressure term. This kinetic theory approach  can used to construct $T^{\mu \nu}$ for ordinary matter (OM) when the geometry is not simple.  This construction of $T^{\mu\nu}$  for OM involves using  averaged equations of motion as well as  equations of continuity for ensembles of particles; it is given in  Section IIA (and in Appendix B).
       
       Next, in Sections IIB \& IIC one focuses  on how to construct   $T^{\mu \nu}$ for DM, incorporating contributions from both $\hat U$ and  $\hat S$ using the model  of DM particles of Paper I.   Two different forms of $T^{\mu \nu}$ are suggested. A criterion for choosing  when either form is applicable is developed in Section IIE\footnote{ One of these forms may actually be that used in cosmology; see Section IID. It can be written more conventionally in the form representing a standard thermal fluid,   provided one adds a cosmological `constant' term $\Lambda$; see \cite{1},\cite{2}. In this approach one finds  $\Lambda$ needs to be variable.}  
       
       Finally,  the representation of a DM halo as an intrusion in a larger DM structure of cosmological size is discussed in Section III and in Appendix C, using one of these two forms of $T^{\mu\nu}$. The outer rotation curves of the large Milky Way galaxy and of the small spiral M33 are successfully modeled by this representation and parameters characterizing the DM halos  derived.  See Figs (1) \& (2).
       
      Most of this paper is `technical', {\it i.e.} it deals with the many details of  setting up a usable physical  description of the properties of the DM gas comprising a halo.  A more casual read would focus on the overall description  provided in the next subsection, 1.2.2, and the ease  which  with many observed rotation curves can be interpreted, given in Section III.  Possible cosmological implications are discussed in Section 2.4, in which it is pointed out that the inclusion of a `$\Lambda$' term may be required if one adopts an inproper equation
      of state for DM.

      \subsubsection{ Comments About  The DM Halo Structure.}
      
           The DM galactic halos, being extended,  are   modeled  in Section III using  an interior Schwarzschild metric  for the halo region. [See eq.({\ref{10})] A typical DM halo  is characterized by a very large momentum flux; this flux formally appears as a `pressure' term when $T^{\mu\nu}$ is written in the conventional fluid form, but it is not to be regarded as a conventional thermodynamic pressure.  The DM  density term is relatively small.  The halos are not in hydrostatic equilibrium. The structure of a halo may be regarded as if it were  a very slowly flowing  interior region  of a  large intergalactic DM cloud in which the galaxy was presently embedded;  details of the flow are determined by conservation of angular momentum.
             
             It was noted in Paper I that using this metric, the motions of non-relativistic particles (not near a singularity)  are really controlled by variations in the metric component, $B$, not by those in $A$.  Accordingly the circular rotational velocity is given by eq.(\ref{C3}) and a large momentum flux $p$, attributable to DM, can mimic `missing mass' when the more conventional interpretation, which disregards the possibility of a momentum flux contribution, is used. For our simple spherical model of the halo region, the variation of $p(r)$  is discussed in Appendix B, Section C and details are provided about how one fits observed rotation curves  with models, using as  representative the  large galaxy, the Milky Way, and  for a much smaller galaxy, M33. [See Section III.]
             
             In general, the halo is divided into two adjoining parts. In the outer halo, some DM particles are not severely constrained by their angular momenta and can `free-fall' towards the central OM galaxy; the pressure and density are given by eqs. (\ref{B13}) In the small inner halo the DM particles' motions are severely constrained by their angular momenta; the pressure and density are given by eqs. (\ref{B14}). [The boundary between the two regions is given by eq. (\ref{B11}).]   For large galaxies, the outer portion dominates and results in `flat' rotation curves. For small galaxies, only the inner halo structure is relevant; this structure results in the `linearly' increasing rotation curves.
     

 \section{ The Construction Of The Energy-Momentum Tensor.}    
 \label{Sec:2}
  {\it  ``...about the dread right-hand side of the Einstein's equations..." }\footnote{Kolb \& Turner 1990, pg 48.}  
 \vskip 6pt

        In most GR models,  for reasons of homogeneity and isotropy, the  source term is assumed to be of the form of the hydrodynamic tensor: 
\begin{equation}\label{3}       
T^{\mu\nu}_{fluid} = (\rho(x) + p(x))\bar U^\mu \bar U^\nu -p(x) g^{\mu\nu}.     
\end{equation}
 Einstein \cite{6} adopted this, regarding   the `pressure' term as representing an averaged momentum flux of the kinetic motions of the OM. In this spirit  we emphasize that the requirement  $T^{\mu\nu}_{\ ;\nu} =0 $   need not be regarded as the equivalent of a thermodynamic equation-of state.  
   In order to connect to  standard  GR models, we shall adopt this ideal fluid  form, eq. (\ref{3}), for representing  the volume average of $ T^{\mu\nu}$ for simple DM models as well, when the local distribution of DM  momentum flux is isotropic. 
   
   The difficulty with eq.(\ref{3}) is that  only in the simplest of ideal situations are the quantities $\rho, \bar U, p$ well defined.   At the very least, one may always add a term
   $\lambda g^{\mu\nu}$ ,where $\lambda=$ constant, to eq.(\ref{3}) effectively modifying the definitions of $\rho \ \& \ p$. If one specifies $T^{\mu\nu}_{fluid}$ accurately, then `in principle' the geometric structure of the LHS of Einstein's can be determined.  The real problem is to reverse the procedure: to `observe'  or adopt a geometric structure and from this to determine the tensor source term. Then one has the problem of interpreting what $\rho, p$ (and the other components of the stress-tensor) mean physically.

    To interpret $T^{\mu\nu}$ for classical OM  one  must start with aggregates of discrete entities. A kinetic theory representation ({\it e.g. } \cite{2}) of them must first be developed and then averages performed  in an enclosure to get a fluid representation.\footnote{The definition of a local enclosure may not be easy; consider the problem of extending the Kerr solution to to include    rotating clouds of particles.  Only in cases involving the simplest time dependences, can co-moving coordinate differences be used to define  local enclosures.}   Such a kinetic treatment for  one constituent  of  OM is given in Appendix B.  It emphasizes that the condition $T^{\mu\nu}_{\ ;\mu}=0$ is equivalent to a strong form of the fluid equation of mass conservation which  effectively includes particle number conservation.

     Normally we regard  $T^{\mu\nu}$ as  a sum of terms, one for each constituent.  Since the LHS of Einstein's equation is non-linear in the affine connections,  $T^{\mu\nu}$ must be {\it inclusive};  in effect the sum of {\it all}   contributions  from individual constituents need to be considered together.  Since one component can contribute most of the mass and another most of the momentum flux, $\rho$ and $p$  for the composite need not be simply related. 
     
      In practice  approximations are introduced. Local mass motions are ignored and $p$ is assumed to be a thermodynamic momentum. This excludes considering  many astronomical problems in which local mass motions transport a momentum flux greater than that associated with a thermodynamic pressure term and one must evaluate the  spacial components of  $T^{\mu\nu}$.   This is the case for the  DM halo models discussed below where the thermodynamic pressure terms are ignored.
    
     In the bold experiment  of constructing a cosmology  for the Universe, we came face-to-face with unknown source terms, normally referred to as `cold dark matter" and `dark energy'.  These  together do not allow a comfortable interpretation of $\rho, p$ as a conventional fluid. The variables $\rho, p, \bar U$ are not really well-defined in physical terms for the unknown source material.  So it should be as no surprise  that the energy momentum tensor for the component representing  DM, discussed below, requires changes in the definitions of $\rho$ and $p$, because the energy and momentum distributions and the equations of motion for DM are quite different from those of OM.

   \subsection{ The Actual Pressure Term In $T^{\mu\nu}$.}\label{BB1}
    First we consider when $p$ is a thermodynamic pressure and when it simply represents a momentum flux for   OM objects.
         For large aggregates of streams of similar particles in a very small region, ${\cal R}$, not near a gravitational singularity,  it is useful to introduce  `number' densities $n_s$, each of which contains a mass factor $m_s$ .  Put $ U_s^\mu = {\bar U}^\mu  + \delta  U_s^\mu$, where
   \begin{equation}\label{B1}
                {\bar N} {\bar U}^\mu = \Sigma_s n_sU_s^\mu,\ \ {\rm  with}\ \ {\bar N} =\Sigma_s n_s\ {\rm and}\  \Sigma_s n_s \delta  U_s^\mu =0. 
   \end{equation}
               Here, ${\bar U}$ and ${\bar N}\equiv \surd (-g) \rho$ are  fields  characterizing the region ${\cal R}$, with $( {\bar N} {\bar U}^\nu )_{; \nu}=0$, giving the ensemble's energy density conservation; set
\begin{equation}\label{B2a}              
 T^{\mu\nu} \equiv \Sigma_s  T^{\mu\nu} _s  = F^{\mu\nu} + P^{\mu\nu},\ \ {\rm where }         
\end{equation}
\begin{equation}\label{B2b}
   F^{\mu\nu} = {\bar N}{\bar U}^\mu{\bar U}^\nu \ \ {\rm and} \ P^{ \mu\nu} = \Sigma_s  n_s\  \delta  U_s^\mu \delta  U_s^\nu 
 \end{equation} 
         with $T^{\mu\nu}_{\ \ ;\nu} =0$.  The trace is $T = g_{\mu\nu}T^{\mu\nu}  ={\bar N}$, the rest-mass density.  The structure equations $T^{\mu\nu}_{\ ;\nu} =0$ are derived in Appendix B, using the individual particles' equations of  motion and the equation of continuity for each stream of particles.
         
         When  ${\cal R}$ is small enough (so that the $\Gamma^\xi_{\mu\nu}$ are slowly varying in it) we assume  the geodesics are nearly constant: $U^\xi_{s; \mu}  \cong 0.$  Now, it is quite remarkable that:
        \begin{equation}\label{B3}
               P^{\mu\nu}_{\ \ ;\xi} =  \Sigma_s   \delta  U_s^\mu \delta  U_s^\nu\  n_{s,\xi} \ .   
         \end{equation}    
         (This follows by differentiation by parts and noting  that because $U^\mu_{s;\xi} =0$, one has $(\delta  U_s^\mu)_{;\xi} = -{\bar U}^\mu_{;\xi}$ )  This means that the local change of $ P^{\mu\nu}$ with position is independent of the affine connections, both frame forces and gravitational forces, the mean velocity ${\bar U}^\mu$, and of details of the variations of  each $\delta  U_s^\mu$ with position.  It is a function only of the stream densities; for this reason we regard it as determined by thermodynamic considerations.  If ${\cal R}$ is embedded in a warmer thermodynamic enclosure, only $P^{\mu\nu}$ is expected to change.

         $P^{ \mu\nu}$  is similar to a conventional kinetic-theory stress tensor.  Under the assumption  of  stationarity, statistical thermodynamic averaging would give $F^{\mu\nu} \to\rho \delta^\mu_0\delta^\nu_0$, and under the assumption of isotropy, for (non-relativistic) streams $   P^{\mu\nu}  \to p \delta^\mu_i \delta^\nu_i, {\rm for} \ i=1,2,3$.  In this simple case the structure equations $T^{\mu\nu}_{\ ; \nu} =0 $ have a thermodynamic interpretation.  However, suppose the region contains part of a rotating stellar disk population;    the spacial components of $F^{\mu\nu}$  would not be zero and the structure equations would not have a  thermodynamic meaning. 
         
         In the simple cold DM halo models developed in this paper,   the variance terms, $P^{ \mu\nu}$, are  assumed very small and ignored.

        \subsection{ The Local $T^{\mu\nu}$ For One Dark Matter Object. }   
        
              There are two  contributions to $T^{\mu\nu}$ from DM objects; see eq (\ref{5}). One is similar to that just noted for ideal OM objects; it corresponds to averaging the particles' transport of mass and translational momentum  along their world lines, $\langle \hat U \rangle$. The other, which we now  discuss, results from the  transport of the particles' internal structure motions. In particular, we focus on the  $\hat S^z$ momentum carried internally by DM particles.                                                                  
        
           As discussed in Paper I, Section II,  {\bf the motions of }a DM object may be described at any point by two vectors $U,S$. Using   local momentum conservation, the only large  internal components contributing  for cold DM
           to $T^{\mu\nu}$ are $\hat U^0 \ \& \ \hat S^z$,  since, in the local frame of reference, $\hat S^0=0$.  We follow the notation of that paper. 
             
              [ One defines $\hat U_s ^\mu =m_0 U_s^\mu \ , \   \hat S_s^\mu=m_1 S_s^\mu $  with $U_s^\mu U_{s \mu} =1,$ time-like, and $S_s^\mu S_{s \mu} =-1,$  space-like.   The index $s$ labels the particular object. Also for halo DM, $m_0=k\Gamma m,\ m_1=\Gamma m$  where $m$ is arbitrary. In the local Lorentz frame of  an object used, $\mu=z$ defines the direction of translation and $S_s^0=0$. We use  multiplicative constants $a,b$ satisfying $a^2m_0\equiv b^2m_1$.  For DM,  $a^2 = \Gamma, \ b^2= k \Gamma$, with $k^2 < 1$ and $\Gamma^{-2} =1-k^2$.  The important parameters $k,\Gamma$ characterize the total space-like internal velocity of a DM object]

          As before in dealing with ordinary matter, we  consider only objects that do not terminate in the volume under study.{\footnote{ This implies $S^\nu_{s;\nu}  =U^\nu_{s;\nu}=0$. See Appendix B, Sections A,B.} To insure  $T^{\mu\nu}_{s ;\nu}=0$ , one needs to  modify the procedure used in Appendix B for OM, because the equations for $ \hat U^\mu,\hat S^\mu$ are not simple geodesics but describe helices. The two equations of motion,  determining the development  of ${\hat U},{ \hat S}$  in a very small region ${\cal R}$ containing their separate paths, effectively differ  only by  multiplicative constants. (See eqs. (14,15 \& 18)  in Paper I. )   Since $T^{\mu\nu}$ is found by summing all streams within a small volume, the distance between  segments of the actual two close-by paths  of the momentum components of  DM object is not relevant. Multiplying  these equations by  constants  to correct for this scale difference, the difference between the two  equations vanishes.  Therefore, we consider as the first part of the internal energy momentum tensor for  a DM   object, comprised of the momentum pair $(\hat U_s,\hat S_s)$,
\begin{equation} \label{4a}          
     \epsilon (a^2+b^2)a^2m_0{\tilde T}_s^{\mu\nu} \equiv a^2 \ {\hat U}_s^\mu{\hat U}_s^\nu -b^2 \ {\hat S}_s^\mu{\hat S}_s^\nu \Rightarrow a^2m_0 \ [m_0U^\mu_s U^\nu_s-m_1S^\mu_s S^\nu_s],          
  \end{equation}           
           where the  $\epsilon =\pm1$; both signs are of  interest. The  normalization  used here is  chosen so that for  $\epsilon =+1$, one has $\tilde T_s \equiv m >0$; this leads to a form useful in cosmology.  By construction  $ {\tilde T}^{\mu\nu}_{s \ ;\nu} \equiv 0$. \footnote{ Using the local Lorentz coordinate system used in discussing the guiding center solution in Paper 1, one finds $ { \tilde T}_s^{xx}={\tilde T}_s^{yy} =0$ so that the `spinning' motion discussed in that paper in the $xy-$plane does not directly contribute to this approximate source term.}    With this form of $T^{\mu\nu}$ no external conservation laws are associated with $\tilde T^{\mu\nu}_{\ ; \nu}=0$. 
           
          For DM  one has $m_0/m_1 =k < 1$  and to good approximation we can exclude the $U^\mu U^\nu$ term when $k$ is sufficiently small.     The dominant term is the one containing $(S^z)^2$.[ We note that for OM carrying small amounts of transverse momentum, one would use $m_1/m_0 = v \ll 1$;  in the limiting case we could exclude the $S^\mu S^\nu$ terms, recovering the form given in equation (\ref{B2a}).]   We assume the internal fields  causing the internal motions represented by the  $U\ \&\ S$ vector fields are included in the mass factors $m_0,m_1,m_2$. The transported components of $\hat S$ are specified at any point by eq.(1) of Paper I;  their  point-to point modification by the presence of  very strong gravitational curvature is not needed or included here.


    \subsection{ An Averaged Local DM Energy-Momentum Tensor.}    
    
       Again, for aggregates of DM particles the momenta transport associated with their averaged motion along their world-lines is specified by  tensors similar to those for ordinary matter, $P^{\mu\nu}, F^{\mu\nu}$  previously discussed; these may be incorporated later in eq.(\ref{5}).   

      For representing the internal motions, we note that In the local Lorentz frames used in Paper I the only  significant stream components contributing to ${\tilde T}_s^{\mu\nu}$ are: $U_s^0,U_s^z,S_s^0,S_s^z$, using the notation of that paper.     For `adding' isotropic ensembles  of similar DM objects  in a small region  to form  volume  averages for the fluid representation we can follow the same procedures used  before\footnote{We can set a particular $S^0_s \equiv 0$ only in one rest frame.  So for an ensemble of DM streams one really get an extra term $\delta \tilde T^{00} =-\bar m_1 {\cal N}\langle (S^0)^2\rangle $ which we will assume is small. From the variance in the distributions of ${U^\mu_s}\ \& \ S^\mu_s$, we get the equivalent of conventional  pressure  terms expressed by  the tensor $P^{\mu\nu}$.  There is only one volume averaging  really involved. That  should include components of the motion associated with the particles' motions around their world-lines. So we may include these in the definitions of $F^{\mu\nu}\ \&\ P^{\mu\nu}$.} in Appendix B for ordinary matter. For $\epsilon =+1$ one finds the leading terms  are:                                        
\begin{equation}\label{4b}
           \langle \tilde T^{00}\rangle   =  + \bar m_0{\cal N}\langle (U^0)^2 \rangle, \ \ \langle \tilde T^{xx}\rangle =   \langle \tilde T^{yy}\rangle =\langle \tilde T^{zz}\rangle \cong -\bar m_1{\cal N}\langle  (S^z)^2 \rangle/3 ,   
 \end{equation}                                                                                                                                                             
   with $\langle \tilde T^{\mu\nu}\rangle =0$ otherwise. Here $ {\bar m}_0,\bar m_1$  are average stream mass  densities\footnote{including the normalization factor  
    $1/(a^2+b^2)$ in the averages.}; ${\cal N} $ is the average number of   DM objects - each represented by a $(U_s,S_s)$ pair-   in a unit volume; and $ \langle(S^z)^2\rangle$ is an average of the squared value of  that component of $S$ which is normal to the spin-plane in the guiding center solution.  [ To simplify we shall consider $\bar m_0,\bar m_1$ as constants with  only the number density ${\cal N}$ satisfying an equation of continuity; see eq.(21). ] Since ${\tilde T}^{\mu\nu}_{s\ ;\nu}=0$ for each object,  one may adopt $\langle \tilde T^{\mu\nu}\rangle_{ ;\nu}=0$ for the average.\footnote{  One may  consider  having many  different ensembles of DM objects in a given volume, each specified by  certain characteristics  {\it e.g.} $m_0$,  energy, etc. In this case,  the quantities listed in eq.(\ref{4b}), including ${\cal N}$, should be labeled by a name; for simplicity, we omit these labels now.}
        
            An ensemble average for the internal motions may be represented by the usual  ideal fluid representation $ T^{\mu\nu}_{fluid}  = (\bar \rho + \bar p) \bar U^\mu \bar U^\nu -\bar p g^{\mu\nu}$, but now    the dominant term $\bar p \equiv \langle \tilde T^{xx} \rangle$ is {\it negative}. This is acceptable because $\bar p$  represents an averaged momentum flux, not a thermodynamic pressure term and other terms contribute to the total energy-momentum tensor.
            
            For the cold DM halo we are interested  in only in the case  $U^0,S^z \cong 1$. (See Paper I.)   Consequently at any two positions $x_1,x_2$ one has  
            $\bar p(x_1) / \bar p(x_2) = {\cal N}(x_1) / {\cal N}(x_2)$; the similar statement for the $\bar \rho$ ratio needs a slight modification. [See eq.(\ref{B8a}).]  
            These relations determine the variation of  the source terms, $\langle \tilde T^{\mu\nu}\rangle$ , expressed as functions of  the fluid variables $\bar \rho$ and $\bar p$.              Because $\bar \rho , \bar p \propto {\cal N} $, one has $\bar p \cong -\bar c^2 \bar \rho/3$ where $\bar c^2 \cong (\bar m_1/\bar m_0) \langle  (S^z)^2 \rangle/ \langle (U^0)^2 \rangle >1$ is to be  assigned or determined for each ensemble of  DM objects; so each ensemble, so defined, is the equivalent of an `isothermal' gas. 
             
               One may try to use this internal energy- momentum tensor as the dominant term in the expression for the total fluid tensor $ T^{\mu\nu}$. [Note this form results from the choice  $\epsilon =+1$.]  Using it we see that the DM fluid would act as to assist compression and resist expansion  because of the negative `pressure' term;  this physics  may be  compensated for by including an additional   source term such as $\lambda g^{\mu\nu}$.  
              
              The alternate form for  ${\tilde T}_s^{\mu\nu}$, resulting from the choice $\epsilon =-1$, turns out to be very useful  for constructing halo models; in this case the components of  $\langle \tilde T^{\mu\nu}\rangle$ are given by eqns.(\ref{4b}) with the signs reversed and similarly for $\bar \rho \ \& \ \bar p$. In both cases, $\bar \rho \ \& \ \bar p$ are specified once  the equation of continuity for ${\cal N} $ is solved; with the assumption $\bar m_0,\bar m_1$ are constants, this is equivalent to solving  $\langle \tilde T^{\mu\nu}\rangle_{; \nu}=0.$

\subsubsection{ The Complete  DM Source Term.}  
            Finally, we  recall we should add in a  conventional term representing the averaged mean time-like momenta fluxes transported through space time along the world-line time-like trajectories, just as we  represented the contribution to $T^{\mu\nu}$ for ordinary matter:
 \begin{equation}\label{5}    
 T^{\mu \nu}_{alt} \equiv \langle \tilde T^{\mu\nu}\rangle +  q_a  \langle V^\mu V^\nu  \rangle,      
 \end{equation}
     where, if $V_s$  represents an averaged $U_s$ motion for a particular DM  object,  $\langle V\rangle $ represents the local volume averaged velocity  for many objects.\footnote{Paper I introduced $V_s$ as a `guiding center' velocity, a mass-weighted averaged component of $U_s$ (without `spin' ) normalized so that $ V^\alpha V_\alpha=1$ and  satisfying $V^\nu V^\mu_{;\nu} =0$. (In a small local volume  one   may average motion along the individual guiding path, determining  mean motions and the averages of conserved quantities.}   We shall assume $q_a = {\cal N } \breve m_0 / \langle (V^0)^2 \rangle$ where $\breve m_0 =\alpha  m_0$ and $\alpha \leq 1$.   When $\langle V^\mu\rangle$ is non-relativistic, it is only the   term $ T^{00}_q \equiv q_a  \langle V^0 V^0  \rangle =\alpha {\cal N }  m_0$ that contributes to $T^{\mu \nu}_{alt}$.  Both tensors in $T^{\mu \nu}_{alt}$
     have  components that vary thermodynamically; in this treatment of the cold DM halos we ignore these terms.\footnote{Also,because we don't know how the DM particles are formed, it is possible that the ratio of the two tensors may not be fixed for any DM aggregate; we ignore this possibility.}
     
     We note that the form  of $T^{\mu\nu}$  for one particle usually adopted for OM is also  composite, consisting both of rest-mass terms and internal motions. See  \cite{YL} .
          
    [ A much simpler approximation results when one can re-interpret the extra $ q_a  \langle V^\mu V^\nu  \rangle$ term as  a small  OM component of the source term associated with the DM component given by eqs(\ref{4b}).  In cosmological models,  we have both DM and OM occurring together  and the precise ratio of one to the other is not known. This extra term can be included in the uncertain specification of  the OM contribution,  allowing us to ignore the problem of specifying $q_a$.]      
          
  
 \subsection{ The  Form  Of $T^{\mu\nu}$ For DM Used In Conventional Cosmology.}     
 
          The form, characterized by $\epsilon =+1$, may be put into the standard  fluid representation of a cold DM fluid used in cosmology by adding an extra term of the form of the `cosmological constant' term to this source term. This  results in  an effective (small) positive pressure term $\bar p_{eff} =\delta p$ in the final fluid representation.
     Let's define 
     \begin{equation}\label{6a}
 T^{\mu\nu}_{dark} \equiv  \ T^{\mu\nu}_{alt} -\lambda(p) g^{\mu\nu}. 
 \end{equation}
     One has that $T^{00}_{dark} =\bar \rho + q_a \langle (V^0)^2\rangle  -\lambda \equiv \hat \rho $ and  for the space-like components {\it e.g.} $T^{zz}_{dark} \cong \bar p +q_a \langle (V^z)^2\rangle +\lambda \equiv \hat p$.    For cold matter, $ \langle (V^0)^2 \rangle \cong 1, \  \langle (V^z)^2 \rangle \ll 1.$ Choose $ \lambda = - p +\delta p >0$ where $\delta p>0$, representing a classical fluid's  positive pressure term, is small; Then $T^{00}_{dark}=\hat\rho \cong \bar \rho-|\bar p| >0$ and {\it e.g.} $T^{zz}_{dark} =\hat  p \cong q_a \langle (V^z)^2\rangle +\delta p =0+\dots$ .  [We may now include the usual pressure and internal energy terms  associated with the neglected term $P^{\mu\nu}$, expressing  the variances in the velocity distributions.]  In the local rest frame all terms of $T^{\mu\nu}_{dark} $  are diagonal and positive and we can therefore write
     \begin{equation}\label{6b}
     T^{\mu\nu}_{dark}= (\hat \rho + \hat p)U^{\mu\nu} -\hat pg^{\mu \nu}  
 \end{equation}
as for a normal fluid.   But then we must rewrite  Einstein's equation as 
\begin{equation}\label{7}
         R^{\mu\nu} -g^{\mu\nu}/2 =8\pi G\  T^{\mu\nu}_{dark} +\Lambda g^{\mu\nu},  
\end{equation}
where we have put  $\Lambda =8 \pi G\lambda$.   It is necessary to  add the extra term  because  $\lambda$ is not a constant but  really varies as $p$ does. Its variability makes no difference because in eq.(\ref{7}) we have added and subtracted the same term to the RHS of the equation.  So we have a `normal'  ordinary matter fluid representation of the dark matter providing we also introduce a {\it variable} cosmological term into Einstein's field equations. Since $T^{\mu\nu}_{dark\ ;\nu}=- g^{\mu\nu} \lambda_{,\nu}$ there is no strict local conservation of energy and momentum for DM  with this source term. We suggest that this is the form implicitly used for representing DM in cosmological models.\footnote{  In the standard   cosmological model $\Lambda$  should be primarily determined by the needed rate of cooling and expansion during the short period of nucleo-genesis; for this  reason
$\Lambda$ probably may be replaced with an average value corresponding to this period. [Even for OM with some spin terms, $p$ is small and not rapidly varying, so a  a small constant value for $\Lambda$ is an acceptable assumption for energies $\leq \sim 100$ Mev.]}   

With this interpretation, setting $\Lambda =\lambda \cong |p|$ and using eq.(\ref{4b}),  one   estimates from observational cosmological parameters a value,  $\langle k\rangle$, of the parameter describing the space-like momentum component  of a typical DM object:
\begin{equation}\label{4c}
 \Omega_{DM}/ \Omega_\Lambda \cong  (\rho-|p|)/|p| = [\bar m_0 \langle (U^0)^2 \rangle  ]/ \ [\bar m_1\langle (S^z)^2 \rangle/3]  -1=3\langle k \rangle-1;    
 \end{equation}
 using   $\Omega_{DM}/ \Omega_\Lambda \sim 1/3$, one finds $\langle k \rangle \sim 0.4 $ for cosmological models.  

\subsection{ A  Local Criterion For $T_s^{\mu\nu}$,  Allowing  The Choice $\epsilon < 0$ for DM. }   

      The augmentation $T^{\mu\nu}_{alt}  \to T^{\mu\nu}_{dark}$ is cosmetic. It was done in order to make $T^{\mu\nu}_{dark}$ look like a familiar fluid stress tensor with positive  density and small positive  pressure terms; the price we pay for this is that we must introduce a `cosmological constant' term in Einstein's equation.  One could use have used the peculiar looking term $T^{\mu\nu}_{alt}$ alone. If we were to drop this cosmetic requirement, the question then arises:  `What additional restraints  must a chosen form for 
      $T^{\mu\nu}$ satisfy in order for it to be acceptable as a source term (without requiring augmentation)?'

             So,  specify a replacement for eq.(\ref{4a}). Perform a thought experiment. Consider an  local ensemble of similar   DM objects  in a local Lorentz frame in which their guiding center motions appear isotropic.   [Equivalently, consider a representative  $(\hat U, \hat S) $  averaged over all possible spacial orientations of the local Lorentz frame.] Then $\langle \tilde T^{\mu\nu}\rangle $ is diagonal  with equal space components in a small region.  Add $\lambda g^{\mu\nu} $ to it where $\lambda = \langle \tilde T^{zz}\rangle$. 
        The resulting sum $T_E^{\mu\nu}$ then has only one component, $T_E^{00}$.   This construct uses Einstein's  original focus on the one term $T^{00} =\rho_{eff}$ as the  source of the attractive gravitational field with the other components of $T^{\mu\nu}$ playing a minor role.
        For example, suppose we used eq.(\ref{4a}) itself; then one finds
\begin{equation}\label{8}        
         T_E^{00} =\epsilon {\cal N} (\bar m_0-\bar m_1/3)  \equiv \rho_{eff}.    
 \end{equation}       
            In our case an attractive gravitational field results with $\epsilon =-1$ for DM if $\bar m_1 > 3\ \bar m_0$ or $ k <1/3$.  After matter decouples from radiation in the standard
            cosmological model, one would expect for DM that $k_s$ would decrease with time.  So it is possible to adopt  $\epsilon =-1$  for the halo models, changing the signs of the  averaged components given in eqs. (\ref{4b}) when $ k$ is small. Halo cold DM will then be charactered by  dominant  positive pressure terms rather than  by the positive density terms characterizing OM.

              [ There is a formal difficulty that $ T \equiv g_{\mu\nu}T^{\mu\nu}$  itself has the wrong sign when $\epsilon =-1$; one would like to match that the sign of the Riemann constant  on the LHS of Einstein's equation . But the procedure outlined determines a value of $\lambda$ and $\lambda g^{\mu\nu} $ may be added to the source terms as a remedy when Einstein's equation is actually being solved.]               
               But, it is always possible to use $\epsilon =+1$ if we introduce the cosmetic $\lambda$ term, so that both $\rho\ \& \ p$ appear positive.  In this case the physical interpretation of the source terms is not direct, but can be done using eqs.(\ref{7},\ref{6b},\ref{6a}).

   \section{Description Of A Simple  DM Halo Structure.}     
\label{Sec:3}
  
        We regard a galactic halo as the {\bf reorganization} of the local intergalactic medium due to the presence of a concentrated  OM gravitational source, similar to the Debye sphere surrounding an ion in a plasma.   Because the gravitational field is small, we can split up the DM into two separate parts and correspondingly consider two separate gravitational fields represented by two different metrics. The  DM halo  part  is superimposed upon a  background  large  intergalactic DM cloud which will represented locally by a fluid with a uniform density $\rho_{00}$ and pressure $p_{00}$.  So, for example, $p_{halo} =p -p_{00}$.   
             
         The DM halo objects move under the attraction of a central galaxy. The local  DM objects  follow orbits modified by their spin terms.  Assuming spherical symmetry for the galactic potential, the DM  orbits satisfy the  integrals of motion given in Paper I. We therefore have conservation of orbital angular momentum for each DM object. Also, the structure equations $T^{\mu\nu}_{\ ;\nu}=0$  are satisfied by conserving the number flux of DM objects. (See Section IIC.)
         
         Our DM halo  $T^{\mu\nu}$ is constructed with $\epsilon =-1$, so that the pressure and density are given by eqs.(\ref{4b}) with the signs reversed.  For convenience we assume little OM in the halo region\footnote{ This has the drawback of ignoring OM galactic disk contributions and should be remedied in future work.} and only one type of low temperature DM, that for which the energetics at $r \geq r_q$ are given by $K: \Gamma(k,0,0,1)$ in a local Cartesian frame.[ See eq.(1), Paper I].  As seen from a point in the halo,  the  attracting central field is weak, $\Psi$, is small.   So $(S^z)^2$ is very much larger than $\Psi$ in the halo and its variation  can be ignored.  
       
  \subsubsection{The  Adopted Structure Metrics.}     
              To represent the local intergalactic DM medium for $r>r_q$ one uses  a Robertson-Walker metric and cosmology, adopting a non-zero cosmological constant to insure   approximate stationarity.  For simplicity, for the galaxy the source contributions of the inner stellar disc population and of the central bulge to the galactic mass are lumped together at the origin.
 For   $r< \sim r_p$  we assume the OM Milky Way galaxy  can be represented by an exterior Schwarzschild metric. For defining the structure of the halo we also assume  spherical  symmetry. The halo, the intermediate region, $r_p \leq r \leq r_q$, is  represented by a  interior Schwarzschild metric   One has for the galactic metrics: 
         \begin{equation}\label{10}
      d\tau^2 =B(r)dt^2 -A(r)dr^2 -r^2 d\theta^2 -r^2 \sin \theta^2d\phi^2,   
     \end{equation}
    where $ A^{-1}\equiv 1-2G{\cal M}(r)/r \equiv 1-2\Phi$. Write     $B\equiv 1-2\Psi$;  its precise form is to be specified by the halo model . For the  exterior Schwarzschild  solution,
    {\it in vacuo}, one has $AB=1$  and  $\Psi=\Phi$.  Details of the representations used  are given in Appendix C.  
    
 \subsubsection{The Local Source Terms}
          For the halo material  consider only  low-velocity DM, for which $V^0 \approx 1$. Adopt eqs.(\ref{4a},\ref{4b})  with $\epsilon =-1$  for representing  the basic DM halo.  In the remainder of this paper, the DM is assumed collisionless and  locally isotropic at large $r$. The  halo source term\footnote{For cold DM  our estimates use  $\langle U^0 \rangle, \langle V^0 \rangle, \langle (S^z)^2 \rangle \simeq 1$.}  may be written as $\bar T^{\mu\nu}  = (\bar \rho + \bar p) \bar U^\mu \bar U^\nu -\bar p g^{\mu\nu}$.  The signs of the components given by eq.(\ref{4b}) are reversed for $\epsilon =-1$. Contributions from the variances of the various velocity distributions are neglected.  Note that the  pressure  term $ \bar p \cong \bar m_1 {\cal N}/3$  is large and  positive with $\bar p \gg |\bar \rho|$.
          
          The  effective halo density $\bar \rho \cong -\bar m_0{\cal N}+q_a $  is  small and  and can be of either sign.  One has a negative  contribution from eq(\ref{4b}) (since 
          $\epsilon=-1$) and  a positive   contribution coming  from the term $q_a \langle (V^0)^2\rangle \sim q_a$ which is hard to estimate without detailed models or constraints on  $q_a\bar m_0$  from observations. To lowest order, since we are already neglecting halo ordinary matter's contributions  we shall assume $\bar \rho \approx 0$. 
                   
         The  evaluation of  how density and pressure in a static halo vary   is done in Appendix B, using  particle flux conservation and angular momentum conservation.  The Appendix is quite detailed because the usual assumption of ``hydrostatic equilibrium" used for OM  does not play any direct role here.The halo structure is given by eqs.(\ref{B13},  \ref{B14}). The DM pressure  terms appearing in $T^{\mu\nu}$  dominate in determining the potential $\Psi(r)$ with variations in the potential $\Phi(r)$, dictated by the average density, playing a very minor role.  It is easy to replicate the observed `flat' rotation curves; the rotation velocity  is given by eq.(\ref{C3}). The formal halo  structure, expressed in terms of the variations  of the metric components,$A,B$ is   outlined in Appendix C.  
         
              \subsection {Two Quite Different Halo Models.}        
                   
         Henceforth, we write $p,\rho$ for $\bar p, \bar \rho$ neglecting halo contributions from OM.  There are two types of  physical halo situations considered. The first, (a), is for a galaxy at `rest' with respect to the local DM intergalactic medium, the halo structure extending out to $r=r_q$.   The second, (b), represents the halo around a rapidly moving galaxy; in this case we expect a bow-shock discontinuity\footnote{ At such a boundary the values of $\rho_a,\ p_a$ would be enhanced over their intergalactic values by the factor  ${\cal N}_a/ {\cal N}_{q}$; using  $V_{gal} \sim \ 200 \ {\rm km \ s}^{-1} $ and $ \hat v_a \sim \ 2-20 \  {\rm km \ s}^{-1} $ for the stagnation velocity as  estimates one sees an enhancement factor of  $\sim 10\ - \ 100$ may occur.} to be traveling in front of the galaxy. Behind the discontinuity we expect a stagnation zone whose  radius $r_a< r_q $ sets the outer halo limit; in this case $p(r_a) =\zeta \ p(r_q) $ where $\zeta \gg1$.                                                                                                            

    In Appendix B, we consider a flux of  incoming DM objects with low angular momenta and show that conservation of angular momentum flux in the outer  halo regions requires 
\begin{equation}\label{9a}   
   p_{ \ halo} \equiv  p-p_{00}= \delta p =[(r_q/r)^2 -1]p_{00} \ \ {\rm  for \ type \ (a) },    
\end{equation}
\begin{equation}\label{9b}
p =(r_a/r)^2p_a  \ \ {\rm for \  type\ (b)}. 
\end{equation}                                                                                                                                                                                             
      
      For type (a), there exists an  $r_a< r_q$ such that for  $r \leq r_a $ the subtractive term term $p_{00}$ can be ignored; then the two equations have the same form for $r \leq r_a$.
 
       The Einstein field equations  for $A\ \& \ B$ (see eq.(\ref{C3})) then directly require that the rotational velocity $ v^2_{cir} \propto pr^2$ for the outer DM halo since there  $\Phi(r)$ is small. Since $p \propto 1/r^2$, these pressure estimates force $v_{cir} \cong$ constant for most of the outer halo where $r_b^\star  < r < r_a$. 
       
          In the very outermost halo region, $r_q\geq r >r_a$ , which we call the `rim', the radial velocity terminates gradually $v_{cir}^2  \propto 4\pi Gp_{00} (r_q^2-r^2)$ for type (a) but more abruptly for type (b), depending upon the width and structure of the putative shock discontinuity.  We suggest adopting $r_q \cong 3 r_a$ for type (a) halos.    
    As one goes inward    the  DM pressure rises to a maximum value and then falls rapidly to zero at $r_p/2$  in a small inner annular zone whose size is determined by the streams'               conserved  initial angular velocity  ($ r_av_a$) and the mass of the central galaxy.  The rather sharp boundary between the zone of freely falling streams, where eqs.(\ref{9a},\ref{9b}) hold,  and the inner zone is given by eq.(\ref{B11}).   We emphasize that eqs. (\ref{9a},\ref{9b} ) do not assume an equation of hydrostatic equilibrium.

    \subsection{  Models For The Milky Way Galaxy and M33.}               
        \subsubsection {The  Observed Rotation Curve.}
        The observational rotation curve for the inner part of the Milky Way, for $R \leq  \sim 8$ kpc is discussed by \cite{9}; we neglect this region.  \cite{10} gives Cepheid observations determining the outer part, $ 5 \leq R \leq \sim 15$ kpc.  The observed outer rotation velocity  curve is quite flat, allowing from noise, perhaps falling from $\sim 230 \pm 15 \ {\rm km \ s}^{-1}$ at $R = 6$ kpc to  $\sim 215 \pm 15 \ {\rm km \ s}^{-1}$ at $R =14$ kpc. We now construct  a DM halo model which  matches these observations.
     
         \subsubsection{ Choosing  Parameters For A Model.}
         
          First, one may predict a rotation curve starting initially with theoretical model parameters.  One may  calculate $p (r),\ \rho(r)$ in the halo  - see Appendix B - and the metric coefficients $A(r),B(r)$; this gives a detailed solution for $v_{circ}(r)$, the rotation curve.  See eq.(\ref{C3}). The mass of the central galaxy ${\cal M}_0$ is needed. Often it is not well estimated, but again this  can be refined by observations.  So only two really arbitrary  parameters, $p_a,  v_a$  need to be specified in constructing a  particular DM halo model.   These two parameters also describe the internal state of the intergalactic DM cloud enclosing the halo.  The choice of $v_a$ is severely constrained by eq.(\ref{B16}). 
          
         The important input data  for such a model halo prediction are  the main `outside' parameters $r_a, p_a$  and an estimate of the average  DM peculiar velocity, $v_a$,  at $r=r_a$.   With this the average intrinsic angular momentum each DM particle carries  at $r=r_a$  can be  estimated.  In practice this quantity can be replaced by the parameter\footnote{ It is defined in the `Approximations' section of Appendix B.}  $\eta$ which , with $R_h$, determine the boundaries of the inner parts of the halo.  Both   $\eta$ and $R_h$, the `inside' parameters, can be estimated from observations  near the `knee' of the rotation curve.   Less accurately, from the `end' of the `flat' rotation curve $r_a$ can be estimated.\footnote{For incomplete rotation curves, the approximation $r_a \simeq 10 R_h$ is reasonable.} Finally the halo density $\rho_a$ at $r=r_a$ needs to be specified; because it is small, it should not significantly affect the structure of the halo. Put $| \rho_a | \equiv  \tilde k_{obs} p_a $ and to first approximation one may put $\tilde k_{obs}(r) \approx 0$.
         
          It is not difficult to reverse the procedure and estimate the model parameters from a given rotation curve by successive approximations.  From the `knee' region one estimates $R_0$ and  sets the potential  due to central OM to be $\Phi(r) =G{\cal M}_0/r  \equiv V_{circ}^2(R_0) \ {\rm at}\ r=R_0$ for $r\geq R_0$; to be consistent ,one provisionally adopts $\tilde k_{obs} \approx 0$. At the far end of the rotation curve  where $r$ is large one has  $p \propto 1/r^2$; therefore  choose a value for $r_a$: then since then  $V_{circ}^2(r) \cong 4\pi G p(r) r^2 \to v_\infty^2$, a constant, one has $p_a$ determined.  Now construct a first approximation to the rotation curve,  $V^2_{first} \approx  \Phi (r) +v_\infty^2$,  calculating from the outside inwards $r \to R_0$. This theoretical approximation will depart from the observations for $r \leq \hat r_h$;  we may take this to determine $\hat r_h \approx r_h$ the maximum compression of the DM; then the observations in the interval $[R_0, r_H]$ determine the value of  $\eta$ (and by establishing the boundary $r_b^\star$, a value for  $v^2_ar_a^2$). 
          
          A further refinement would include a contribution to $\Phi (r)$ from the OM disk population for $R_0 \leq r \leq \hat r_h$; this could be determined for the Milky Way from the Gaia observations. With this known, non-zero values of $\tilde k_{obs}(r)$ may be investigated

\subsubsection{ Parameters For The Milky Way.} 
           
        As an example, we constructed a  representation for the Milky Way galaxy's rotation  curve for $r \geq 6$ kpc, using a type $(b)$ model.   For the OM galaxy, we adopt  $V(r)^2 =G{\cal M}_0/r$, with $V(r= 6 \ {\rm kpc})   =220\ {\rm km\ s}^{-1}$.  This model  uses $\eta=0.5$ and $R_h=8$ kpc as the effective  inner edge of the dense part of the  DM halo; no DM can go below $\sim 4 $ kpc and the boundary between the inner and outer halo zones is at $r_b^\star =12$ kpc. Eq. (\ref{C3}) gives  $v_{cir}(r)$. 
        
        We use the   approximations\footnote {These  greatly simplify  the the detailed halo calculations in the range $r= 4-7$ kpc  where the halo contributions are very small; they are good approximations in the range $r=9-12$ kpc. }   discussed in Appendix B; in this case one sets $v_{cir}^2 \to v_\infty^2 =4\pi G(p_a+\rho_a)r_a^2$ for $r \gg R_h$ for the asymptotic halo solution. We chose $v_\infty = 175 \ {\rm km\ s}^{-1}$ and $\rho_a \approx 0$. 
 The   outer `edge' of the halo   was chosen to be   $r_a \approx 60-80$ kpc; this corresponds this results in estimating $v_a \approx 30-25 \ {\rm km \ s}^{-1}$. using eq.(\ref{B16}).  The results are given in Table 1.

 \begin{table}[hbt]
\caption{ 
Model $V_{cir}$ vs. r For Milky Way}
\label{tab:my-table}
\begin{tabular}{ | l | l | | c | l |}
\hline
6 kpc  & 237 km/s & 14 kpc                & 227 km/s \\ \hline
7      & 230      & 16                    & 221      \\ \hline
8      & 227      & 20                    & 212      \\ \hline
9      & 227      & 40                    & 195      \\ \hline
10     & 228      & 60                  & 188      \\ \hline
12 kpc & 234 km/s & 80 kpc & 185 km/s \\ \hline
\end{tabular}
\end{table}
  
  
  \begin{figure}
  \includegraphics[scale=0.5]{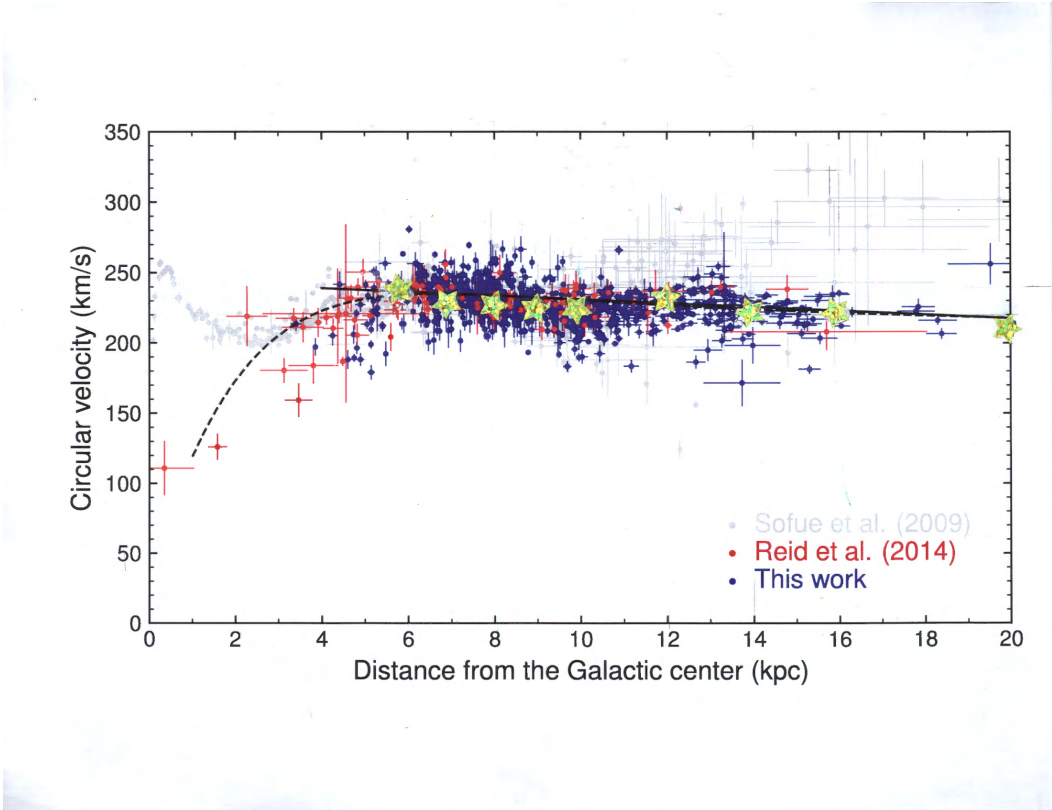}

  \caption{Fig.1  The observed rotation curve for the Milky Way, given by  \cite{10} . The calculated model points of Table 1, represented by gold stars, are superimposed on their results.}  
  
  \end{figure}
  
       This is in very good agreement  with the results of \cite{10}; see their Fig 2  and their mean solid line.   [Our tabulated values are  perhaps $\sim 5 \ {\rm km\ s}^{-1}$ too low.]  For data covering from $6 -16$ kpc , they find| \footnote{Another of their models gives $V_{cir} = 
            \approx 222\ {\rm km\ s}^{-1}$ at  $r=7.6$ kpc .}$V_{cir} \approx 234\ {\rm km\ s}^{-1}$ at  $r=8$ kpc and a linear fall-off, $d V_{cir} /dr =-1.34 \pm 0.20\ {\rm km\ s}^{-1}/{\rm kpc}$. This Cepheid data is in good agreement with the maser data \cite{10b}. The observations of $v_{cir}$ for $r> 14$ kpc still have large error.   
             
            [Caveats: For the inner Milky Way , a spatially averaged rotation curve is not available and it is known that some inner regions on opposite sides of the galaxy exhibit differing rotation curves  \cite{11}.  Also, estimates of the luminous mass ${\cal M}_0(r)$ differ by a factor of two  \cite{10c,12,13,14},  reflecting differences in reduction procedures. Another fit  to the data used in the rotation curve of Clemens \cite{9} would  allow a lower value $R_h \approx 6$ kpc;  this probably the lowest value  acceptable. Within the uncertainties of interpretation of the observations, somewhat larger values of $R_h$  up to $9-9.5$ kpc are not really excluded. Small variations in $\tilde k$ are permissible; We used $ \tilde k \equiv  | \rho |/p=0$.   For $\tilde k =0.1$ , $V_{cir}$ is reduced by $ \sim 2 \ {\rm km \ s}^{-1}$ at $r=12$ kpc.]
            
   \subsubsection{The Rotation Curve for M33}  
   
     If the central galaxy hasn't enough mass, or the angular momentum of the dark matter in the halo is sufficiently high, then there may not be a   ``free-fall" outer zone of the halo .  Then the dark matter concentration  in the vicinity of the galaxy is  severely limited by its conservation of angular momentum, the pressure being given by its ``inner zone" value, see eq. (\ref {B14}). But because the small spiral M33 has only $\sim 1/10$ the mass of the Milky Way  galaxy in its  interior,  even a small contribution from a dark matter accumulation in its vicinity leads to a rise in the rotation curve. We assume the observed region corresponds only to the inner DM halo region.

     As with the Milky Way , the motions in the interior of the small spiral, M33 are quite complicated \cite{KCC}. In the region  beyond $\sim 15$ arc minutes, $\sim 3$ kpc, the rotation curve seems reasonably represented by the observations of \cite{CS}.  We adopt $v_{cir} =95 -100\  {\rm km \ s}^{-1}$ at $r= 15$ arc minutes as characterizing  the $1/r$ potential representing the OM central galaxy.  Also take the outermost points of the rotation curve as determining $r_h \cong 80$ arc minutes. with $v_{cir}(r_h)=145\ {\rm k \ s}^{-1}$.  Then the only unknown model parameter is $\eta$; its choice is specified by
     matching the observations in the entire region $15 \leq r \leq 140$ arc minutes.  One finds the assignment $\eta = 0.15\ - \ 0.18$ reasonable.

   \begin{figure}
   \includegraphics[scale =0.5]{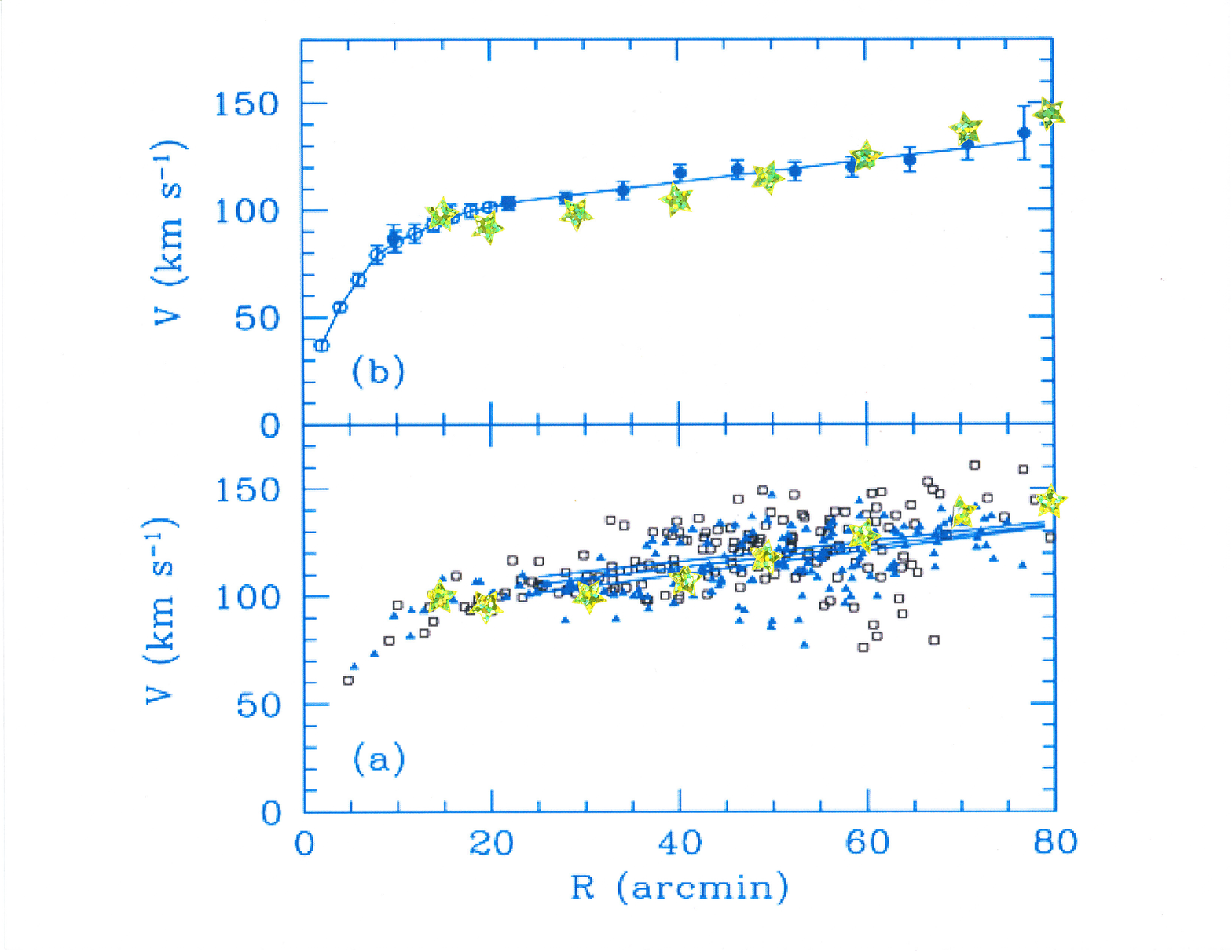}
    \caption{The observed rotation curve for the outer portion of M33, as given  by \cite{CS}.  The model points, represented by gold stars, are superimposed upon the observations in (a), and   compared to the linear representation some people prefer, in (b). The model shown used $\eta =0.15$.  Here, 5 arcmin $\approx 1$ kpc.}
    \end{figure}
    
       \subsubsection{   The Model Parameters Seem Acceptable.}   
    
      In addition to the requirement that the model predicts   observed rotation curves, one requires that the model parameters  are reasonable.  For the Milky Way, arbitrarily assuming $r_a =10R_h \sim 80$ kpc, one finds $p_a/c^2 \cong 6.0 \pm 0.3  \times 10^{-27}  {\rm g \ cm}^{-3},$  the `errors' reflecting   uncertainty  in the choice of $\tilde k_{obs}$.  Alternatively, using\footnote{ The mean peculiar speed of DM objects in originating regions is designated by $v(r_a) \equiv v_a$. In paper I it is suggested that the mean intrinsic rotational speed of a DM object is in equilibrium with  its peculiar translational velocities.  This numerical estimate assumes the DM objects have baryonic masses.}    $v_a \sim 30 \ {\rm km \ s}^{-1}$ and  $ r_a^2v_a^2\equiv \ G{\cal M}_0R_h$ with $ {\cal M}_0 \approx 10^{11}{\cal M}_\odot$  \cite{10}, one estimates $r_a=63  \pm 15\ {\rm kpc}$; the uncertainty reflecting the uncertainties in $R_h$ and ${\cal M}_0$.   The probable range for the DM pressure is $p_a/c^2 \sim 1.5 \times 10^{-26} \ {\rm g \ cm}^{-3}$  at   $r_a\sim 50 \ {\rm kpc}$  to $p_a/c^2 \sim 0.5 \times 10^{-26} \ {\rm g \ cm}^{-3}$ at   $r_a\sim 80 \ {\rm kpc}$ .
   
     For the large spirals, the extent of the  observed haloes sets a lower limit  to the value of $r_a$. Using $r_a \sim 50 -100$ kpc, this range for $p_a/c^2$  needed looks reasonable.  For both types (a),(b) of halos  it is reasonable to adopt $p_a \sim 10\  p_q$ where $p_q$ is the effective density of DM in intergalactic space.  These should be compared to other appropriate  densities (See \cite{15,16}.)The limiting cosmological critical density is $\rho_c \sim 10^{-29}\ {\rm g \ cm}^{-3}$. The mean densities of luminous matter in clusters of galaxies are $\sim  10^{-26} -10^{-28} \ {\rm g \ cm}^{-3}$;  this is on the order of our estimate for $p_q/c^2 $.               
  
     In the MW inner halo the DM pressure  is $p_b=p_a(r_a/R_0)^2 \approx 50-100\ p_a$ corresponding to  $p_b/c^2 \sim 10^{-24}-10^{-25}\  {\rm g \ cm}^{-3}$ which is slightly smaller than the OM  mean density of the MW galaxy distributed inside a sphere of  $10$ kpc radius,   $\langle \rho \rangle \sim 2 \times 10^{-24} \ {\rm g \ cm}^{-3}$. The local mean density of OM   is estimated to  be $\sim 1.4 \pm 0.6 \times 10^{-23} \ {\rm g \ cm}^{-3}$ \cite{10}.
     
     For the representative small galaxy M33, we may estimate $p_b$ at $r_b \approx 16$ kpc from $v_{circ}^2 \cong 4\pi G p_br_b^2$. Because of the observed  linearity of the rotation curve the value of $p_b$ is reasonably determined  even though the value of $r_b \cong r_h$ is uncertain. One finds $p_b/c^2 \approx 1.1 \times 10^{-25} \ {\rm g \ cm}^{-3}$.
     The agreement of this value  of $p_b$ with the  value found for the large Milky Way galaxy suggests one `infall' mechanism is sufficient to explain both types of rotation curves.

   \subsubsection{Improvement Of The Modelling Procedure.}  
   
     To significantly improve these models one needs observations to estimate the amounts of ordinary matter  entrapped within the DM haloes and flowing through the inner edge  of the DM halo.  Also, the mass of  the OM disc population in the outer regions, $r \geq \eta R_h$ is needed. It is possible that outer OM disk populations may cause `wiggles' in the rotation curve. With a combined OM and DM model one could determine the  density contribution  $q_a$ of eq.(\ref{5}). 
   
    This is important because  the only model parameter not yet well determined is $| \rho_a | \ \equiv \tilde k_{obs} p_a$.  Trials in fitting indicate $\tilde k_{obs} $ is small, so to lowest approximation we used $\tilde k_{obs} \approx 0$.  To determine it from the rotation curves requires assumptions about the outer OM galactic disk density and the subsequent modification of $\Phi(r)$.  We do not have a theoretical estimate of this parameter.  [See the discussion of  eq.(\ref{5}).] 
          
  \subsection{ Observational Challenges.}      
       
          A main  observational challenge is to determine the interface between the outer halo and the intergalactic medium $r_a \leq r \leq r_q$ and the value of $r_a$. Observations of DM lensing \cite {21} in earlier epochs are modeled with $r_q >100$ kpc. In the case of the MW galaxy,, which exhibits a  flat rotation curve at $\sim 200\ {\rm km \ s}^{-1}$, one has limits set by  disturbances at $\sim 60$ kpc   by the  dwarf galaxies LMC \& SMC, and at $\sim 700$ kpc by M31; evidently the assumption of a spherical gravitation potential would would be crude and useful only in selected directions. A study of MW `halo objects' suggests the halo changes character at $\sim 30$ kpc and might extend to $\sim 120 $ kpc  [\cite{4}]. Also very low surface brightness observations  of  emissions from ordinary matter trapped in the  extended DM regions around galaxies would be rewarding in possibly mapping out motions in the halo.
     
     It should be remarkably easy to use this procedure to represent the observed halo radial velocity curves of many galaxies. The derived model parameters may be used  as a first probe  of the variations in the intergalactic  DM medium.  There  really is an interesting unexplored observational field here. There is no reason to assume that since the epoch of decoupling from ordinary matter, most of the DM  (- $\sim 3/4 $ of the total matter -)  should remain anywhere near homogenously distributed in the Universe.

     Not all galaxies enveloped in a DM intergalactic medium need have  haloes such as those calculated. Our model requires that the peculiar velocities of a significant portion of the local DM be low. The components of DM carries intrinsic angular momentum which, combined with  orbital angular momentum,  forms  a centrifugal barrier to prevent close approach to the center of attraction. (See Paper I.) The  DM halo's inner edge effectively starts at $r_p \cong R_h\equiv r_a^2v_a^2/G{\cal M}_0$.  A low mass galaxy may have its DM halo starting well beyond its visible structure. A high mass galaxy may have its halo structure begin inside its  visible structure. The Milky Way  probably is such a galaxy.   So far, observations of other large spirals suggest  values $R_h \sim 4-8$ kpc.  Any OM entrained with the DM at  large distances, having only orbital angular momentum will fall through the DM barrier at $r_p$.  However  one guesses it should be a minor halo component;  one expects OM to constitute $\sim \Omega_b/\Omega_d \sim 0.2$ \cite{7,8} of the halo density at large $r$.

     \subsubsection{Other  Considerations.}

                 We also note that DM with high angular momentum may provide the external pressure boundary condition  needed to construct  classic Bonner-Ebert    models  of protogalaxies. \cite{17,18} Because of past galaxy-galaxy collisions, it would be impossible to rule out the presence of DM inside  galactic disks.  DM entrapped within galaxies by galaxy-galaxy collisions may thermalize and loose  almost all its orbital angular momentum, thereby providing a confining pressure to aid form very massive star  models.
                 
                   DM with larger peculiar  velocities than those   assumed  ($ \sim 30\  {\rm km \ s}^{-1}$) could form  haloes with similar structures but at much larger distances; this may provide models of  galaxy lenses.  We have not allowed for the presence of OM in the haloes. The presence of the Sunaev-Zeldovich  effect, indicating the presence of OM , precludes  direct application  of these models to the haloes of galaxy clusters.

   \section{ Summary.}

     A  model of the  dark matter (DM) surrounding galaxies has been developed which represents the observed rotation curves and can explain the absence of DM in  small potential wells such as the solar neighborhood. The model is based on conservation of angular momentum in the flow of DM around a galaxy.  In it (see Paper I) the DM is treated as a new state of matter in which matter while traveling along time-like paths also carries considerable transverse momentum. This transverse momentum results in  intrinsic angular momentum terms. These, coupled to the usual orbital angular momentum,  form centrifugal barriers, preventing  DM objects from passing close to centers of gravitational attraction.  In addition, the DM objects can transport an internal form of momentum. For aggregates of moving DM objects, the local sum of the internally  transported  momentum fluxes appears as a new form of `pressure', not a thermodynamic pressure. For DM this term is large and must be included in specifying  the energy-momentum tensor, the local  source of the gravitational field and in calculating rotational velocities in the outer parts of galaxies.  The halo cannot be primarily OM: (1)  it is not seen; (2) there is no reason why
     OM halo material would be diminished in the region $r_p \leq r \leq r_h$.
     
     \subsection{Halo Structure}

     Two alternate forms of the energy-momentum tensor representing  DM are derived in this paper.  One  features a  large positive momentum flux appearing as a dominant pressure-like term.  Using   this the structure equations  for the halo are developed  and examined;  these show that the pressure  increases considerably as one approaches the central attractor but  suffers an abruptly termination  because of the centrifugal barrier.   The DM matter forms an annulus around the central galaxy, a `halo', and does not enter the central region.
 
 In this halo region, the circular velocity is given by one of Einstein's equations eq.(\ref{C3}). Here $\Phi$ represents (mainly) the central OM galaxy's gravitational potential and is small and slowly varying.  In the outer halo regions $p \propto 1/r^2$, so $v_{cir} \cong $ constant.  The variation of $p$ is determined by consideration of particle flux and angular momentum conservation.
 
   A model for the Milky Way galaxy uses an  inner edge of the halo at $ R_h = 8\ {\rm kpc}$ and an outer edge at $ r_a\sim 80 \ {\rm kpc}$;  the  pressure there is $p_a/c^2 \sim 6 \times 10^{-27}\ {\rm g\ cm}^{-3}$, with the intergalactic DM pressure being perhaps a factor of 10 lower. The local intergalactic DM contributing to the MW halo 
   seems to have low peculiar velocities $v_{pec}\sim 30\ {\rm km \ s}^{-1}$. A model fitting the rotation curve of the  small spiral M33 is constructed using only the inner structure of the halo; in it the centrifugal barrier  is important.
 
     Aggregates of such DM objects  could comprise the main component of the intergalactic medium.  Because of the gravitational attraction of an (ordinary matter) galaxy, the intergalactic DM medium locally compresses to form a galactic halo around a galaxy.  One expects, because  of variations in the local peculiar velocities of the DM objects, that the size of the halos would vary. 
  Models of galactic haloes can be calibrated from observations  with sufficient accuracy to  provide a tool for investigating  variations in  density,  pressure and  the peculiar velocities of the intergalactic DM medium.  

  \subsection{Cosmology}   
     The other form of $T^{\mu\nu}$  for DM found  can be made to resemble the conventional form of  $T^{\mu\nu}$ used in cosmology to represent ordinary matter (OM), providing  a cosmological `constant' term $\Lambda$, slightly varying, is also introduced.  This raises the possibility that at least some part of the `dark energy'  is not a separate entity but due to the representation of DM used. 
    
    If the alternate representation  $T^{\mu\nu}_{dark}$ is adopted for cosmology,  then  $\Lambda \sim 8\pi G p$ is not constant and the Standard Cosmological Model  would need some re-discussion.  We do not claim this is the only reason for the $\Lambda$ term's presence  in Einstein's equation, but this simple DM proposal provides a rationale for its presence.  It is needed in modern cosmology to get the observed baryon acoustic oscillation spectrum and was based on the assumption that the DM was simply like an undetected form of ordinary matter with an ordinary fluid $T^{\mu\nu}$  featuring a conventional $p -\ \rho$ relation.   However, this spectrum represents only the epochs when the Universe was becoming transparent , when the optical depth is less than $\sim 3$ and does not justify the assumption that $\Lambda$ was constant at earlier epochs.
         
      It is possible that $k =|\rho |/p$ varies on cosmological time scales.  The observed cosmological parameters  \cite{8}  $\Omega_d/\Omega_\Lambda \approx 0.15/h^2 \sim 0.3$   would then represent an averaged $k$; it would very likely represent the value of $k$ when particle formation was frozen, {\it i.e.} when the temperature was $\sim 1-5$ Mev.

\vskip 6pt                     
               [(An earlier version of some of these ideas is available on-line \cite{22}.]
          
 \section{Appendices.} 
 \section{ Appendix A:  Astronomy.}    
\subsection{Inferred   Properties Of Dark Matter.}
        There exist many large spiral galaxies (including the Milky Way) with `dark matter' haloes.  The observational characteristic of these spiral galaxy halos are galaxy rotation curves, $v_{rot}(r)$ consisting of: (1) a central spherical contribution followed by a (roughly) linear part :    $v_{rot}(r) = V_0(r/r_{00})$ for $r \le r_{00} \sim 2\ -\ 4$ kpc;  (2) a (fairly) `smooth' transition zone $r_{00} \leq r \leq r_0 \sim 4- 8$ kpc;  and (3) a  roughly ``flat" part $v_{rot}(r) = V_0$ for $r_0 < r \le r_1 \sim 16\ - \ 50+$ kpc. This outer limit is  hard to estimate and in a few cases may be $ \geq 100 $ kpc. A  value $V_0 \sim 200\ {\rm km \ s}^{-1}$ characterizes large spirals.  \cite{K1,K2,K3,K4,K5,K6}.  For the Milky Way see \cite{10}.\cite{10b}.
       These rotation velocities\footnote{These are defined by the observed  line-of-sight motions of extreme   Population I objects such as very luminous HII regions or  HI gas and molecular clouds, known to depart by less than $\sim 5 \%$ from  circular orbital velocities $v_{cir}(r)$ in galactic disks outside the very central region. The absence of these objects in the outer parts of galaxies limits our present knowledge of the rotation curves at large $r$. } are normally interpreted as circular velocities, setting $v_{cir}^2 =G{\cal M}_{obs}/r$  where ${\cal M}_{obs} \propto r$ when $r_1 \geq r \geq r_0$ and regarding  ${\cal M}_{obs}$ as a true mass distribution.  No galactic mass distribution compatible with the observed stellar distributions predicts this behavior in the outer zone \cite{11}.  An unseen (`dark') matter component is  usually hypothesized, probably amounting to several times the mass contribution  inferred for the stellar and interstellar matter contribution. Because the  halos are strongly concentrated to the center, the   halo DM must be moving at non-relativistic velocities; this is also true for DM in clusters of galaxies:     $v < \ \sim 400 \ {\rm to} \ < \sim10^3\ {\rm km \ s}^{-1}$ (in clusters) are conservative upper limits.  
       
       There is also a class of smaller spiral galaxies showing smaller halos whose outer rotation curves have an approximately constant positive  slope.  The  small galaxy M33 studied in this paper -see Fig.2- may be considered a prototype.

        Presently, there is no  direct evidence  that  this `dark' matter (DM)  interacts with ordinary matter (OM) through electromagnetic interactions or collisions.  However, in the standard cosmological model  both types of matter remained coupled to radiation until $T\sim 1$eV. To be coupled the DM  must have an interaction cross-section with $n(t)\sigma cR(t)/\dot R(t)>1$. Allowing for the model's uncertainties we note that for DM  to remain coupled for energies  down to $T \sim10$ eV (or $1$ KeV) one must have an interaction cross-section $\sigma >10^{-29}\ ({\rm or} \ 10^{-33})\ {\rm cm}^2$, if DM is represented by particles of baryonic mass.  [If they have  less than an electron's mass, one has the additional problem of explaining their low velocities in galactic halos.] Also, such particles  would be entrapped within a typical star like the sun if $\sigma > 2 \times 10^{-37}\ {\rm cm}^2$.  


                
\section{  Appendix B: Supplementary Physical   Arguments.}   
  \subsection{ The Vanishing Of The   `Line' Divergence  $U^\nu_{s;\  \nu}$} \label{A1}                         
          The technical point here is that we are are considering functions defined only along a line so we cannot directly use the Gauss divergence theorem with no sources or sinks present.  However $U^\nu_{s; \nu}$ is  a scalar, independent of a coordinate system choice. At any point, choose a geodesic coordinate system and rotate the coordinate axes so that $\tau$ is the coordinate along one of the axes. Then $U_s^\mu$ has only component, $\mu=\tau$, and   $dU_s^\mu/d\tau =0$ because $U_s^\tau$ is  a  geodesic's tangent vector.
          For DM when we consider confluences of streams such as $S_s $ in a small region, we shall also assume $S^\mu_{s;\mu} =0$ to prevent local creation or destruction of streams; rigorously, in the presence of spin one only has $\langle S_s^\mu \rangle_{;\mu} =0$ established in the guiding center approximation.
  
 \subsection{ $T^{\mu\nu}$ For Ordinary Matter.}    
                                                                                                                                                                                                                             
    We start with a method for constructing
  $T^{\mu\nu}$ for  ordinary matter (OM) when mass motions are significant and their associated momentum flux exceeds any assignable pressure term.  An example is the rotating stellar disk in a spiral galaxy.  It may be useful for numerical solutions to Einstein's equations
   
  The only formal requirements in GR on the fluid energy-momentum  tensor are that it is symmetric $T^{\mu\nu}=T^{\nu\mu}$ and divergence-free
  $T^{\mu\nu}_{\ ;\nu} =0$; consequently, a term $\lambda g^{\mu\nu}$, where $\lambda$ is constant, can always be added to  $T^{\mu\nu}$.  
        
      For each  OM  object $s$ with constant mass $m_s$ following the  time-like path $x^\mu_s(\tau)$ we define an energy momentum tensor by using the geodesic equation 
      $U^\nu_sU^\mu_{;\nu}=0$ and requiring  mass conservation. Set
 \begin{equation}\label{1a}    
      T_s^{\mu\nu} \equiv m_s f_sU^\mu_sU^\nu_s   
      \end{equation}
            along the path and determine the variation of $f_s$ by requiring
  \begin{equation}\label{1b}
                   (m_sf_s U^\nu)_{;\nu}  \equiv 0   = [m_s f_s \surd (-g)U^\nu]_{, \nu}\ \ \ {\rm or}      
   \end{equation}          
    \begin{equation}\label{1c}      
   d \hat m_s/d\tau_s + \hat m_s U^\nu_{s,  \nu} =0, \ \  { \rm where} \ \  \hat m_s \equiv m_s \surd (-g) f_s ,   
   \end{equation}
then  $T^{\mu\nu}_{s \ ; \nu}  = 0$ and $T^{\mu\nu}_s $ can be used as a fluid's source term for determining  the curvature tensor. So, for satisfactorily representing even one typical object as a source, one must introduce a `density' distribution function. Then the distributed object may be thought of as equivalent to a localized stream with a finite cross-section and mass density $\hat m_s$.
                                                                                                                                                            
     One may show (see  Appendix A)  that $U^\nu_{s;\  \nu} =0$ if the objects are neither created nor destroyed in the region under study.\footnote{This divergence-free condition is equivalent to choosing the radiation gauge  in EM theory.}  Henceforth we drop the accent, writing $m$ for $\hat m_s$ when speaking of a `typical' object. Then  eq.(\ref{1b}) has three useful forms depending upon the functional form of $m$ available: (1) steady state one-dimensional pipe flow\footnote{ {\it i.e.} For an aggregate of close-by objects all moving similarly,  the argument of Appendix A1 holds and we can regard them  collectively as a narrow confined stream of objects.}: 
 \begin{equation}\label{2a}    
    m v ={\rm constant};     
 \end{equation}                                                                                                 
(2)  given the {\it fluid} form $m_s(\tau)=m_s( x(\tau))$, the  equation of mass continuity results,
\begin{equation}\label{2b}
      \partial_t \ m +   {\vec v} \cdot \nabla m=0,  
  \end{equation}
 where $U^\mu_s =\Gamma (1,{\vec v})$; and (3), given the {\it kinetic} form $m_s(\tau)=m(\ x(\tau), U^\mu(x(\tau))\ )$, Liouville's equation results, 
 \begin{equation}\label{2c}
 [U^\mu\partial_\mu -\Gamma^\mu_{\alpha\beta }U^\alpha U^\beta (\partial/\partial U^\mu] m=0,    
\end{equation}
 which plays an  important role in interpreting some stellar kinematics. \cite{3,4}  Each of these equations hold only on a time-like path $x^\mu(\tau)$. 
                              
   Eqs.(\ref {2b},\ref{2c}) use an additional constraint,  $\nabla \cdot \bar U=0$, for  individual  particles and for simple\footnote{ We  may need to allow for the exchange of energy and momentum among different components of the source terms.} particle ensembles.   It is needed when the total number of particles is conserved.  For a fluid this concept has no meaning, so it must be imposed on fluid representations to enforce the particulate nature of matter.  
   
   [Depending upon the interval of space-time being studied, this condition can be somewhat relaxed if the total source term   $T^{\mu\nu}_{fluid}$  is broken down into a sum of constituent fluids. For example, consider  radioactive decay of a nucleus $A \to B +C +\gamma$.  The number densities $n_A, n_B,n_C$ are modified, also their associated momenta are changed and 4-momentum is transferred to the radiation field tensor.. The fluid $T^{\mu\nu}$, representing each species, is changed, but their sum in the source term  in GR is not.  So, {\it e.g.}  $T^{\mu\nu}_{A ;\nu} \neq 0$. But because the source term $T^{\mu\nu}_{fluid} $ is inclusive we can use the ideal fluid form $T^{\mu\nu}_{A ;\nu} = 0,$  and ignore interactions such as particle destruction. Similarly for considering star densities and the formation of black holes and proto-stars. ]                 
 
  Now, suppose we have a finite number of streams in a small  region ${\cal R}$.  Conventionally we treat  ${\cal R}$  as an energy-momentum reservoir and effectively define local {\it fields }  $\bar N\bar m_s$, $\bar U$ and $\bar T^{\mu\nu}$ extending throughout  ${\cal R}$ as averaged values of the included streams.   [The averaging is  detailed in text.] Then  $\bar T^{\mu\nu}= F^{\mu\nu} + P^{\mu\nu}$ where  only $P^{\mu\nu}$, expressing contributions from the variances of the velocity distributions,  need require a thermodynamic equation-of-state to evaluate; its contribution is usually small for non-relativistic OM.  

  We emphasize that the spacial components of $F^{\mu\nu} $  represent the streams' average momentum flux  \cite{5}, and not really thermodynamic `pressure' terms.  The assumption of isotropy in cosmology allows confusion between the components of  $F^{\mu\nu}$ and of  $P^{\mu\nu}$; but in modeling individual galaxies it is the streaming motions of various galactic populations   contributing to $F^{\mu\nu} $ that are the source of the spacial terms in the stress tensor.    
                                                                                                                                                   

\subsection{ The Halo Structure.}\label{B1}                                                                                                                

     \subsubsection{ Representation Of The Averaged $T^{\mu\nu}$ In The DM Halo Model.}\label{BB2}  
          
       We use only DM with $\epsilon =-1$, so that  the components of the basic energy momentum tensor are given by eq. (\ref{4b}), with the signs reversed to represent the source function.   From  eq.(\ref {5}) we must add a contribution from the other source term $q_a\langle V^\mu V^\nu \rangle$; for cold DM only the (positive) density term contributes. So, for  the sum of the two contributions to the halo density we write  $\rho = -\tilde k p$, where $\tilde k$ is expected to be small.
       
               First we derive eqs.(\ref {4a}, \ref{4b}), slightly generalizing the notation to allow  inclusion of anisotropic distributions of DM.
             In the  halo annular region ({\it e.g.} $r_{\hat p} \le r  \leq r_q$ in the spherically symmetric case)   specify the halo $T^{\mu\nu}$ by the change of  $T^{\mu\nu}$ from its value in the background model (at  {\it e.g.} $r= r_q$).  The halo material responds to the gravitational attraction of the central attractor, and the DM with sufficiently small angular momentum is compressed as it falls in.      In type (b) halos DM  falls inward from a stagnation zone on time-like trajectories at nearly parabolic\footnote{ In the integral of energy conservation along a typical guiding center path, eq.(20) of Paper I, ${\cal E} \cong 0$.}speeds as judged by an observer at $r_{\hat p}$; {\it i.e.}  $v(r_a) \ll \surd (2 \Phi(r_p) \approx \surd (2G{\cal M}/r_p) $. In general we consider low-temperature halos in which at $r\geq r_a$, the same is true and most of the DM objects have small peculiar velocities at distance and are isotropic.
  
   We  write  for the change in a typical  spacial diagonal component  of $\delta T^{\mu\nu}$ at the  halo point  $r\geq r_p$, 
   \begin{equation}\label{B4} 
   \delta p ={\cal N}mP -{\cal N}_q m_qP_q= \left[{{\cal N}\over {\cal N}_a}{m\over m_a}{ P\over P_a} -\zeta \right] p_{aa}.  
   \end{equation}
   Here ${ m}$ is an average mass density, ${\cal N}$ is the number of DM objects in a unit volume and ${ P}$ represents a  local value of one spacial component  of $\langle  T^{\mu\nu}\rangle$. We consider only the major  velocity contributors to these averages. For the isotropic case $P$  represents the square of the `speed'\footnote{ Alternatively, this corresponds to  $s^2\equiv (S^r)^2   +  J_s^2/r^2 $ transported locally along the mean `guiding center 'paths $x^\mu_s (\tau_s)$ discussed in Paper I, and refers to that component of $S^\mu$  not involved in the internal  `spin' motion.} $ \langle(S^z)^2\rangle/3$  of eq.(\ref{4b}).      Here, $p_{aa} \equiv \ {\cal N}_a\ m_aP_a $ and $\zeta p_{aa} ={\cal N}_q m_qP_q$, so that $\zeta $ is small for $r \ll r_a$ and one may write $\delta p \cong  p.$ [A similar expression holds for $\delta T^{00}$ in terms of the average value of  $(U^0)^2$.]  Again, ${\cal N}$ need not  specify   all the local DM, but just a  particular subset.
      
   Using the steady state one dimensional pipe flow, (see Sec. II) one has $  m(r) /  m_a = s_a/s(r)$ where $s$ is the speed along the fluid's streamline .   By eq(1) of Paper I,  $S^z  \sim 1$ is  very large compared to $\Phi$, and we may take it not very much affected by the gravitational field in the halo and we can set 
     $m/m_a =S^z_a/S^z \simeq 1 -\Phi_a+\Phi \sim1 \ \ {\rm and}\ \ P/Pa =(S^z)^2 /(S_a^z)^2 \simeq 1-2\Phi+2\Phi_a \sim 1$. Hence for  halo DM
     \begin{equation}\label{B5}
      p /p_a ={\cal N}/{\cal N}_a.               
 \end{equation}
 The density term, $T^{00}$, can be expressed  in terms of $p/p_a$ and the potential $\Phi$ in eqs.(\ref{B8a}, \ref{B8b}).

    For non-spherically symmetric geometries, the rule to follow is that the vector component $S^z_s$ is transported along a particle's world-line with the  velocity $u^z_s$. So, if in a local spacial Cartesian coordinate system, $\grave x, \grave y, \grave z$  one has for the ensemble's mass motion tensor $F^{\mu\nu}$  only the diagonal elements $a_i^2 (\grave U^i)^2, \ i=1,2,3$, one should use $P \to T^{ii} =a_i^2 (S^z)^2/3, \ i =1,2,3.$

     \subsubsection{Momentum Flux.}                                                                       
        The pressure terms in eq. (\ref{4b})  represent  a type of momentum flux.  Statistical treatment of large numbers of  streams  of DM carrying momentum flux can be modeled after  the  standard treatment of  radiative flux  to introduce a momentum density. (See {\it e.g.} \cite{19,20}.)  Normally one writes for the energy (or momentum)   crossing a surface element $dA$ in a solid angle $d\Omega$ oriented at an angle $\hat \theta$ to the normal of $dA$ in a time interval $dt$,
\begin{equation}\label{B6}
        dF=(I/v\pi )\  v\cos\hat  \theta \  d\Omega dA dt,     
\end{equation}            
               defining $I \equiv I(\theta,\phi)$, the local specific intensity; it may be labeled by other parameters, such as the energy.  Here $v$ is a  velocity along the axis of the solid angle. The standard procedure is to focus on the development of $I$, rather than $F$, since it represents a `density'.  One has {\it e.g.}  $(I/v\pi ) = {\cal N} \langle h\nu\rangle$, the energy density in isotropic radiation flows. For a steady-state when $I$ is independent of $\theta,\phi$,  the total transfer of flux crossing $dA$ in one direction  in unit time is $dF=  I dA$.  For us, dealing with momentum flux,   $I \propto {\cal N}$ of eqs. (\ref {B4}, \ref{B5}), the local density of streams.    For this momentum flux, $ I \propto {\cal N} \langle (mS^z)S^z \rangle =p$, using $v=S^z$.  So to determine $p$ we focus on determining the flux of DM streams in a halo.  For isotropic flows through concentric spherical annuli   in free space one requires    for  flux conservation:
\begin{equation}\label{B7}
            4\pi r_a^2 I(r_a)=4\pi r^2 I(r) \ \ \ { \rm or} \ \ r_a^2/r^2 = {\cal N}(r)/{\cal N}(r_a)=p(r)/p(r_a), 
  \end {equation}          
               For the mass carried by the incoming streams one  takes $I =\langle\bar  m_0 \rangle {\cal N} \cong \langle m_0U^z\rangle {\cal N}/\langle U^z\rangle$; since conservation of mass gives $\langle m_0U^z\rangle=$constant along a stream-line, one  has,
  \begin{equation}\label{B8a}             
     \rho(r)/\rho(r_a) \cong \langle U^z(r_a)/U^z(r)\rangle\  p(r)/p(r_a); 
  \end{equation}  
               here $U^z$ is that component of $U^\mu$ perpendicular to the  `spin' plane, $ (U^z)^2 \equiv\langle U^r\rangle^2 + \bar L_v^2/r^2$.  (See eq. (20) of Part I.)  In the outer  part of the halo $\langle U^z(r_a)/U^z(r)\rangle \cong 1$.  In the inner part, one has $\langle U^z\rangle$ determined  by the gravitational potential (see Appendix C):
\begin{equation}\label{B8b}
             \rho(r)/\rho(r_a) \cong [\Psi(r_a)/\Psi(r)]^{0.5} p(r)/p(r_a) . 
 \end {equation}        
   \subsubsection{   Angular Momentum Conservation. }      
   
               Now examine momentum conservation for a particular aggregate,  with number density ${\cal N}$, of  incoming DM objects which at large $r$ have specified  low angular momenta;  ${\cal N}$, considered as a function of $r$,  is subject to the constraint that each object  in the aggregate must conserve its associated angular momentum in a spherical potential; the gas is collisionless.         
               
                Consider flow through a portions of a sphere; the normals of the area elements are the radius vectors.    For convenience suppose  at some large $r = r_a$,  in a volume element all the streams  of objects considered  have the same stream density $\langle m_a\rangle$ and the same (small)  average speed $v_a$   (where $U_s^\mu =\gamma(1, \vec v_{a,s})$).

  For each stream of objects the orbital angular velocity $L_s=r_a \times (r_a U^\phi_s) = v_a  \ r_a  \sin \theta_s \equiv v_a b_{a,s} $ is  a constant and is determined just by the value of   $\theta_s $ ( the angle between the 3-vectors $r_a$ and $\vec v_{a,s}$). The total number of {\it inflowing} streams  at  a shell of radius $r_a$  is then $ F_a={1\over 2}\cdot 4\pi r_a^2 \cdot \int I_a \cos \theta \ \psi(\theta, \phi) \sin \theta d\theta d\phi$, where $\psi$ is a distribution function for the ingoing streams.  For large $r$ we adopt a distribution $\psi=1$, corresponding to velocities being isotropic\footnote{The specification of $\psi$ is part of the specification  of the angular velocity distribution; in principal it could be a function of $v_ar_a$.  In near equilibrium the  symmetries of the  velocity distribution reflect the symmetries of the effective gravitational potential;  see  \cite{3} .} so $ F_a ={1 \over 4} \cdot 4 \pi r_a^2 I_a$;  the collection of all  these streams carries orbital angular momenta up to $L_a=r_a v_a$.  
                                                                                                                                                                                   
    Let us follow such an aggregate moving in towards the center, assuming for convenience that $v_a$ is small enough that each stream follows a nearly parabolic orbit, {\it i.e.} $v_a^2 \ll  V_0^2(R_0)$, where $V_0(R_0)$ is close to the galaxy's maximum rotation speed. For  any large $r=\tilde r< r_a $, a similar argument gives ${\tilde F} =\pi \tilde r^2\tilde I$ for the  inflowing streams.  There is a region of `free-fall' in which the streams are not appreciably deflected;  in it  $\tilde F =F_a$ so  eqn. (\ref{B7}) is satisfied.
    
\subsubsection{The Inner Halo}    

       Inside this region streams with large  orbital angular velocity $L_s$ will be deflected. We calculate the loss to the inflow.  In the background space suppose we regard each sphere  represented by a (fictitious) collection of  `bound' DM circular orbits, each with an associated orbital angular velocity given by $L_{cir}^2=r_{cir}^2 v_{cir}^2(r)$.  For a falling stream to pass through a sphere of radius $r_{cir}$ it must have its distance of closest approach to the center be less than  $r_{cir}$. 
       
        For example, let $\Phi =G{\cal M} /r \equiv V^2(r)$ with ${\cal M}$ a constant, representing the (ordinary matter) inner part of a galaxy for $r \geq R_0$.  For a stream of DM particles following a parabolic orbit (${\cal E}\cong 0$) one requires   $L^2_s \leq 2L_{cir}^2  +W_u^2 \equiv L_\ast^2(r) $  where $W_u$ represents the intrinsic angular momentum associated with the DM object (See Paper 1.)   One has $L_{cir }^2 =G{\cal M}[r_{cir} -r_p]$ where 
        $r_p \equiv W^2_u/(G{\cal M})$.  [ Bound circular orbits cannot be defined for DM  for $r_{cir} \leq r_p.$] With parabolic infall velocities, DM objects have their turning point at $r=r_p/2$.

       Consider a particular sphere $ r_b\leq r_a.$ Only those streams with low $L_s$ reach $r_b$; they have the same (invariant) values of $L_s$ as they had when they were at $r=r_a$.  We calculate the fraction $f$ of inflowing streams at $r_a$  that can reach this particular $r_b$; one has $f =(r_a v_a)^{-2}\int_0^{L_\ast} LdL $, using  $L=r_a v_a \sin \theta$ as the variable of integration instead of $\theta$:
      \begin{equation}\label{B9}
        f \cong r_b(r_b-r_p/2 ) V^2(r_b)/(r_a^2v_a^2). 
               \end{equation}

        This loss must be compensated for by an decrease in $I_b$  since the total number of streams  with assigned $L_s$ which can travel from $r_a$ to $r_b$ is conserved.   Consequently\footnote{ Note we have not assumed that at our starting point, $r=r_a$, $v_a r_a$ is the maximum orbital angular velocity possible, $v_a^2 \leq v_{cir}^2(r=r_a)$.} since $\pi r_b^2I_b = f \cdot \pi r_a^2I_a$, one has, for an inner zone
    \begin{equation}\label{B10}   
       I_b/I_a ={\cal N}_b/{\cal N}_a =  p_b/p_a =\tilde v_b^2/ v_a^2 \equiv V^2(r_b) [1-r_p/(2r_b)]/v_a^2;   
      \end{equation} 
this definies $\tilde v(r_b)$.      This holds\footnote{ More general potentials can be used.  We really assume that   in the outer halo $r v_{cir}$  is an increasing function of $r$, so that there is a limiting radius $r_b^\ast $ at which `free-fall' ends and for which $r<r_b^\ast$  equation (\ref{B10}) holds.}  for all $r_b \leq r_b^\star$ for which $  \tilde v_b r_b \leq  r_a v_a$.  

\subsubsection{The Boundary Between The Two Halo Parts.}

 The boundary between the two regions, $f=1$, occurs at $r=r_b^\star$ where
 \begin{equation}\label{B11}
  r_b^\star -{r_p \over 2} \cong v_a^2 r_a^2/(G{\cal M}_0 ); 
  \end{equation}   
 the RHS assumes $r_b^\star$  is close to the inner halo edge  where  $G{\cal M}(r) \simeq G{\cal M}_0 \ =r_b^\star V^2(r^\star_b)$.  This is true for only high mass galaxies and DM with low initial angular velocities, $r_av_a$;  this is the case of most interest. Then the DM  generally is in the free-fall zone where $p \propto  r^{-2}$. But, for low-mass galaxies $r_b^\star$ may be close to $r_a$,  severely limiting any `free-fall' zone.
  
  In this  inner zone  the change in the gravitational potential can be appreciable because of contributions from OM galactic disk populations and $\rho$ actually has a complicated behavior; by itself eq. (\ref{B8b}) gives
  \begin{equation}\label{B12}
  \rho(r)| \propto   p(r)/\surd \Psi(r).   
\end{equation}

\subsubsection{  The Imput Parameter $v_a$.}
     From the discussion in Section 3.2 in text one has  for the Milky Way galavy, $r_b^\star \simeq 12$ kpc.  Using the approximation eq.(\ref{B11}) and $r_a \approx 10 r_h \cong 80$ kpc one estimates $v_a \sim 25\ {\rm km \ s}^{-1}$. Considering the uncertainty in $r_a$, this is in good agreement with its estimate in text, based on the observed value of $v_\infty$.

 For the small galaxy M33, the estimate is more uncertain because less is observed and $v_\infty$ is not known. The text uses $r_h \cong 16$ kpc and $r_b^\star \cong 18$ kpc for representing the rotation curve. For theoretical reasons, one expects $r_b^\star$ to be `close' to $r_a$.  If we use $r_a \approx r_b^\star$ then  eq.(\ref{B11}) inplies $v_a \sim 40  {\rm km \ s}^{-1}$.  If $r_a \approx  2 r_b^\star$, then $v_a \sim 20  {\rm km \ s}^{-1}$.    Also, because of the natural width of the distribution of $p$ DM near its peak, $r_h$, it is possible that a lower value for $r_h$ should have been inferred:  $r_h \cong 12 - 14 $ kpc , resulting in $\sim 20 \%$ increase in $v_a$. 
 
   It is interesting, then, that both galaxies  can be characterized by the same imput velocity, $v_a \sim 20 -40  {\rm km \ s}^{-1}$ despite their differences in size and OM mass.

\subsubsection{ Some Caveats.}

          The infall argument holds for ordinary matter as well as dark matter since it involves conservation of a flux, which may be a mass flux as well as a momentum flux. If some of the in-falling streams do not come from a real stagnant region and actually have appreciable mean  stream velocities at large distances, the treatment would need modification, since  the expression for $f$ holds only in the stream's rest frame.  In this paper, using our expression for $f$, we are actually considering a restricted class of incoming matter and our estimates for densities, etc. at large distances refer only to them.  Because of collisions, ordinary matter particles can exchange angular momentum and some part of the initial flow will always reach the central galaxy with very low angular momentum and fall in. But, as postulated, the DM  particles always possess a non-zero total orbital angular momentum and cannot penetrate deeply into the central galaxy.

  \subsubsection{  Approximations For The Halo Pressure \& Density.}

  For the region $r \leq r_a,$  use as unit of length  $R_h \equiv r_a^2v_a^2 /(G{\cal M}_0) $ ( which shall turn out to be close to the maximum of the rotation curve $R_0$ for many cases of interest).   Represent the intrinsic angular velocity by
  $W_u^2= r_a^2 (\xi v_a)^2$ with the maximum orbital angular velocity of the incoming DM halo streams given by $L_u^2 =r_a^2v_a^2$.   For  the form of $T^{\mu\nu}$ adopted, $\rho\leq 0$.  f One writes $\rho_a=-\tilde k p_a$ and estimates $\tilde k \leq 0.1 $ but in this section we allow for the possibility of a much larger value.
 The formal inner edge of the halo, defined in terms of the lowest  bound `circular'
 path is $r_p =\xi^2 R_h$ and the furthest inward DM parabolic orbits can go is $r_p/2 =(\xi^2/2)R_h \equiv \eta R_h$;  we will use $\eta \approx {1\over 2}$  as representative in the approximations because we are interested only in infall with low angular momentum.  In these units  the boundaries of the inner zone are given by $\eta R_h \leq r \leq (1+\eta)R_h =r_b^\star$.   We used  $\eta =1/2$ in our models.
   
     In the outer region one has, from eqs.(\ref{B7},\ref{B8a}):
     \begin{equation}\label{B13}
  p(r)=p_a (r_a/r)^2;\ \ \rho(r) / \rho_a \simeq p(r)/p_a ,  
  \end{equation}                                                                                      
 ignoring the variation of  the small quantity $\Psi$ in the outer zone in calculating $\rho$.
  
   In the inner zone, using eqs. (\ref{B8a},\ref{B10}) we allow for the variation of $\Psi$ in calculating the density. Using the notation $p_b \equiv p_a (r_a/r_b^\star)^2; \ \rho_b \equiv \rho_a(r_a/r_b^\star)^2$ and $r=xR_h$ (so that $\eta \leq x \leq 1+\eta$ in the inner zone ),          
 \begin{equation}\label{B14}
  p  =(1+\eta)^2p_b (x^{-1} -\eta x^{-2} ) ; \ \ \rho=(1+\eta)^2 \rho_b [ (x/(1+\eta)]^{0.5}(x^{-1} -\eta x^{-2} ). 
  \end{equation}
For the DM mass in the inner zone, using $M_1 =(4\pi/3)\rho_bR_h^3 <0$, one has
 \begin{equation}\label{B15a}
 \delta {\cal M}_1(r) \equiv \int_{\eta R_h}^r 4\pi \rho r^2 dr = M_1(1+\eta )^{3/2} \ [ 6x^{5/2}-10\eta x^{3/2}  +4\eta^{5/2}]/5, 
 \end{equation}
 so that, using $\eta =0.5$  one has $\delta {\cal M}_1(r_b^\star) \cong 3M_1$. Also, for  the additional DM mass above the boundary, $r_b^\star$, in the outer zone, one finds
 \begin{equation}\label{B15b}
 \delta{\cal M}_2(r) \equiv \int_{r_b^\star}^r 4\pi \rho r^2 dr= 3(1+\eta)^2 M_1[x-(1+\eta)] . 
 \end{equation}
 In the inner zone the pressure reaches a maximum at $r=R_h$ and then, going inwards falls rapidly to zero at $r=\eta R_h$. One finds the halo structure effectively starts at $r \cong R_h$, {\it i.e.} for $r_p/2 \leq r <R_h$, $p,\rho$ may be taken as small. For $r_b^\star \geq  r \geq R_h$,  one may consider the  pressure and density as approximately constant; they  then fall $\propto 1/r^2$ as we go further out.   For either type (a) or type (b) halos the size of the halo, using $ R_hR_0 \approx R_0^2$,  is constrained  by the  angular momentum condition
 \begin{equation}\label {B16}
  r_a^2 /R_0^2   \cong V_0^2/v_a^2 ,     
\end{equation}
 where $V_0^2 \equiv G{\cal M}_0/R_0$; this imposes an important constraint on the choice of $v_a$.
         With these expressions for $p,\rho$ the structure of the halo can be developed  and the functions $A(r), B(r)$, used in defining the metric, can be calculated. See Appendix C.
 
 \subsubsection{ The `Shielding' Distance  $r_s$}
 
        For type (a) halos, which may be quite extended, there is an additional feature because the model  DM halo density may be negative if $q_a$ is small. In the limiting case   one may define a `shielding' distance, $r_s >r_b^\star$, such that
        \begin{equation}\label{B17}
       \delta {\cal M}_1(r_b^\star)+\delta{\cal M}_2(r_s ) +{\cal M}_0 =0, 
       \end{equation}
      since $\rho <0$ for DM. This corresponds to $\Phi(r_s)=0$ and in Appendix C is used to define the edge of the halo $r_q$.   Since $r_q$ is For $r < r_q$   Using $\eta =0.5$ and   $\rho(r_b^\star) \equiv-\tilde kp_b$, and for the outer zone $p_b(r_b^\star)^2 =p_ar_a^2$. one finds  using equation (\ref{C3}) for the observed asymptotic circular velocity, $v_{\infty}$, that 
      \begin{equation}\label{B18}
       r_s/R_h \cong (1.5\tilde k) V^2(R_h)/v_\infty^2,  
       \end{equation}
   where $ V^2(R_h) R_h \equiv G{\cal M}_0 $.  For the MW galaxy $r_s/R_h \approx 2.2/ \tilde k$.   The significance of $r_s$ is that for $r>r_s$ the details of the potential representing
   the central attractor are no longer relevant and the metric coefficients are purely determined by the background intergalactic DM cloud.  It is reasonable to consider $r_s \geq r_q>r_a $ for type (a) halos and this the proper definition of $r_q$.  Again, this  value of $r_s$ is crude because the value of $q_a$  has been assumed low and we have neglected positive density contributions from OM trapped in the halo.

\section{ Appendix C:  Metric Parameters For  A Galaxy's  DM Halo.}    

      The DM halo is represented by the standard fluid energy momentum tensor $T^{\mu\nu}= (p+\rho)U^\mu U^\nu -pg^{\mu\nu}$  (with $c^2=1$).    When $r_p \leq r \leq r_a \leq r_q$ one has $p$ as a large pressure term  and $ \rho <0$  with$ |\rho | \ll p$. For $r$ outside this range,  $T^{\mu\nu}_{halo} \cong 0.$
 
    \subsection{Outside}                                                 
  The intergalactic medium is represented by the Robertson-Walker metric
  \begin{equation}\label{C1a}
  d\tau^2 =d \bar t^2 -R(\bar t)[ (1-\bar kr^2)^{-1} dr^2 +r^2d\theta^2 +r^2 \sin^2 \theta d\phi^2]   
  \end{equation}
   where the structure equations are:
   \begin{equation}\label{C1b}
    3\ddot{ R} =-4\pi G(\rho + 3p -2\lambda)R, \ \ \dot{ R}^2 +\bar k =8\pi G (\rho +\lambda) R^2)/3              
    \end{equation}
   and  the `energy' conservation constraint 
   \begin{equation}\label{C1c}
   3(p+\rho) d \ln R/ d \bar t =-d(\rho +\lambda) /d \bar t        
   \end{equation}
    hold for $r>r_q$.  Here,  we have added $\lambda g^{\mu\nu}$ to the usual fluid source term. One can get a static model  (with $R=1$) by choosing $\lambda$ appropiately.   For our case of the  surrounding intergalactic cloud  being purely DM, $p \to p_{00}$ is the dominant term with $\rho \to \rho_{00}$ being small and negative. For a quasi-static model   take $\rho + 3p -2\lambda =0$ so that $\lambda \cong  3p_{00}/2$ . Also $\rho+ \lambda \cong \lambda$ so that one may take $\dot R \cong 0.$
    Then  $\bar k \cong  4\pi G p_{00}$,  having the dimensions of an inverse length squared, sets the scale of the curvature effects. For example, for
     $ p_{00}/c^2 = 10^{-27} {\rm g\ cm}^{-3}$ one has $  1/\surd \bar k \equiv L\simeq 10^9$ l.y. Consequently, one may take $\bar k \cong 0$ for our halo models.

 subsection {The Form Of $A(r)$ In The Halo.}          

    In the Schwarzschild model (equation (6) of Paper I))  for the entire halo $r_p/2 \le r \le r_q$, we have $A=1/(1-2G{\cal M}(r) /r)\equiv1/(1-2\Phi)$ where ${\cal M}(r) ={\cal M}_0 +\delta {\cal M}$; ${\cal M}_0 $ is the mass of the central attractor  at $r \approx 0$ and $ \delta {\cal M}$ is given by eqns. (\ref{B15a}, \ref{B15b}); For DM , $ \delta {\cal M}<0.$  The halo begins at $\eta R_h$ with $\eta \sim 1/2$; there one has $\rho =p =0$.   Eqs.(\ref{B13}, \ \ref{B14}) give approximations for $\rho, p$ for the halo.  Using these approximations for the mass variations   one has
      \begin{equation}\label{C1}
   \Phi (r) =G {\cal M}_0 /r +  4\pi G \rho_b r^2 (1- R_h^3/r^3 ) /3 ,\ {\rm for}\ \  R_h \leq r \leq r_b^\star,   
   \end{equation}                          
       where $r_b^\star =R_h(1+\eta)$.  This matches  There is a similar expression for $\Phi (r)$ for $ r_b^\star \leq r \leq r_a$, resulting in {\it e.g.}
       \begin{equation}\label{C2}
         \Phi(r_a) = G( {\cal M}_0+{\cal M}_1)/r_a+ 4\pi G\rho_ar_a^2[1-r_b^\star/r_a],    
         \end{equation}
    where $\rho_ar_a^2 =\rho_b (r_b^\star)^2$ has been used.  One has $v_a^2r_a^2 =G{\cal M}_0 R_h$ and, for massive galaxies $[1-r_b^\star/r_a] \approx 1$.  [See eq. (\ref {B11}.] 
     As noted in Appendix B, there is a limit to $r_a$, provided by a shielding length $r_s$, see eq.(\ref{B18}), if the (negative) halo density provides significant shielding. To join onto the exterior solution, theRobinson-Walker metric, at    $r=r_a$ one choose the parameter $\bar k$ appropriately. 
             
\subsection{The Form of $B(r)$  In The  Halo}         

        The restriction $ T^{\mu\nu}_{\ ;\nu}=0$ requires $\Psi^ \prime \cong { p}^\prime/({\rho} +  p)$. But this standard relation is not directly useful because: (1) $ p,  \rho$ have different dependences on $\Psi$; (2) there are other neglected contributions, see eq.(\ref{5}); and  (3)$\Psi$ has an additional contribution from the OM central galaxy.  Also, the quantity $ k \equiv - \rho / p$ may not be constant in the inner halo. In the outer parts of DM halos , the variation of $ p$ is predictable and $\Psi$, mainly due to DM,  can be directly determined from observations  (since $v^2_{cir} \equiv -r\Psi^\prime$) assuming strict circular motion.  
     Generally   other combinations of Einstein's equations are more useful for determining approximations for the three quantities, $ p,  \rho,\Psi$. Because $ k$ is small  and its variations should have minor effects in the outer halo, we  initially assume it  constant.

     For the Schwarzschild metric,  we have $A^\prime /A+B^\prime/B =8\pi G(p +\rho)r A$. Again using $B=1-2\Psi$, where $\Psi$ is small in a halo model, one may rewrite this {\bf Einstein equation} in a more useful form:
     \begin{equation}\label{C3}
     v_{cir}^2 \equiv -r d\Psi/dr = 4\pi Gpr^2 + \Phi(r),  
     \end{equation}
     where $\Phi$ is known -see the previous section- and for most of the halo is small. For our simplified version of $T^{\mu\nu}$ for DM one has $ \delta {\cal M} <0.$
      Both $p$ and $\rho$ are continuous at the joining points $r=r_p/2$ and $r=r_b^\star$.  Using  our approximations for the density, eqns.({\ref{B13},\ \ref{B14}) for $ r_p/2 \leq r \leq R_h$, one has $\Psi(r) \cong \Phi(r)=G {\cal M}_0 /r$ in the  innermost interior halo zone.  Next one has
      \begin{equation}\label{C3a}
     v_{cir}^2 = 4 \pi G p_br^2 +G {\cal M}_0 /r +4\pi G \rho_b r^2[1-R_h^3/r^3]/3\ \ {\rm for}\ R_h \leq r < r_b^\star. 
     \end{equation}
     
     For the outer zone, since $p \propto 1/r^2$, one has
     \begin{equation}\label{C3b}
      v_{cir}^2 = 4 \pi G p_ar_a^2  +4\pi G \rho_a r_a^2[1-r_b^\star/r] +G [{\cal M}_0+{\cal M}_1] /r  
      \end{equation}
    for $ r_b^\star \leq r < r_a$.  Again $\rho_a r_a^2 \cong \rho_b(r_b^\star)^2$ , and $\rho_a =-kp_a, {\cal M}_1 < 0$ [if there are no contributions from OM]. For large $r$, one has
    $v_{cir} \cong $ constant.
    
  \subsubsection{ The Outer Rim of the Halo.}    
              
             In  the  outermost halo region, $ r_q\geq r>r_a$   where $r_a$ is large and $\delta \rho = -k\delta p$ (see eq.(\ref{9a})) one has
             \begin{equation}\label{C4a}
              \Phi(r) =  G {\cal M}_T/r +4\pi G(-k  p _{00})\ [r_q^2(r-r_a)-(r^3 -r_a^3)/3] /r \    
              \end{equation} 
              with 
              \begin{equation}\label{C4aa}
                  \Phi(r_a) =G {\cal M}_T/r_a \ {\rm and} \  \Phi(r_q) =G {\cal M}_T/r_q  -8\pi G kp_{00} r_q^2 /3.       
              \end{equation}
              Here $ {\cal M}_T=  {\cal M}_0 +\int_{R_h}^{r_a} 4\pi \rho(r) r^2 dr  $.  In practice we  use $r_q \sim 3 r_a$.  Then, to satisfy the boundary condition $A \to 1$
              at $r=r_q $   we suggest   adding the small constant $\Phi_0 \equiv  -\Phi(r_q)$ to $\Phi(r)$ throughout the entire region  $r \leq r_q$.\footnote{ For the usual exterior Schwarzschild  solution representing an isolated source , one chooses $\Phi_0 =0$ because the solution is supposed to hold as $r \to \infty$. Here we impose another boundary condition.} We shall ignore it for $r \leq \sim r_a$ 
              although it may affect estimates of the shielding distance $r_s$.              
                Using eqs.(\ref{C3}, \ref{9a}), one has
              \begin{equation}\label{C3c} 
              v_{cir}^2 = 4\pi G p_{00} [r_q^2 -r^2] + \Phi(r) -\Phi_0  
              \end{equation}
                  This shows  both $v_{cir}, \ \Psi^\prime \to 0$ as $r \to r_q$.  Since we only determine $\Psi^\prime (r)$,  see eq.(\ref{C3}), a constant $\Psi_0$ may be added to it.  We suggest requiring $\Psi_0 =-\Psi(r_q)$ to satisfy the boundary condition $B \to 1$. Then $d \bar t^2 \to dt^2[1-2\Psi(r)]$ establishes the clock rate comparison inside   and outside; this is to be expected because the halo is in a gravitational potential well with the halo model sitting on top of the background DM cloud model.


\begin{thebibliography}{}                     
                          
 \bibitem[Helfer(2019)]{H1} Helfer, H.~L. 2019, \apj  880, 73 . Paper I
  \bibitem[Yang \& Liu(2018)]{YL} Yang, Y.~B. \& Liu, Key-Fei, 2018 \prl 121, 12001 
 \bibitem[Kolb \& Turner(1990)]{1}  Kolb,E.~W. \&  Turner, M.~S.  1990, The Early Universe (NY, Addison Wesley) 
  \bibitem[Weinberg(1972)]{2} Weinberg, S. 1972,  Gravitation and Cosmology( New York, Wiley)   
   \bibitem[Chandrasekhar(1942)]{3} Chandrasekhar,S. 1942  Principles of Stellar Dynamics (U. of Chicago:Chicago)
      \bibitem[Battagia et al(2005)]{4} Battagia, G. et al 2005 MNRAS 364,433
      \bibitem[Landau \& Lifshitz(1959)] {5}Landau,L. ~P. \& Lifshitz, E.~M. 1959, Fluid Mechanics (Pergamon: Oxford) 
    \bibitem[Einstein(1950)]{6} Einstein,A. 1950, The Meaning Of Relativity, 5th ed. (Princeton:Princeton)
     \bibitem[Tegmark et al (2004)]{7} Tegmark, M. et al 2004 astro-ph/0310723v2 
     \bibitem[Longair(2005)]{8} Longair, M.~S. 2005, in  IAU Symposium 201 ``New Cosmological Data and The Values Of The Fundamental Parameters" , ed. by  
                    A. Lasenby and A. Wilkinson (Sheridan Books: Ann Arbor)
      \bibitem[Clemens(1985)]{9}Clemens, D.~P. 1985 \apj 295, 442                                        
     \bibitem[ Mr\'{o}z et al(2019)]{10}Mr\'{o}z, P. , Udalski, A.,  Skowron, D.M., Skowron, J. Soszy\'{n}ski, I., Pietrukowicz, P. Szyma\'{n}ski, M.K. ,Poleski, R., Kozlowski, S. 
     and Ulaczyk, K.2019 \apj  870 L10
  \bibitem[Read et al (2014)]{10b} Read, M.~J.,Menten,K.~M., Brunthaler,A. et al   2014\apj 783,130 
  \bibitem[Trimble(2000)]{10c} Trimble, V. 2000 in  Allen's Astrophysical Quantities, 4th ed.,edit. byA.N.Cox( Springer:NY)
\bibitem[Mihalas \& Binney(1981)]{11} Mihalas, D \& Binney, J. 1981 Galactic Astronomy, 2nd ed. (W.H. Freeman:San Francisco) 
  \bibitem[Bahcall et al(1983)]{12} Bahcall, J.~N., Schmidt, M., \& Soneira, R.~M. 1983 \apj 265, 730
  \bibitem[Bahcall(1986)]{13} Bahcall,J.~N. 1986 ARAA 24,577   
   \bibitem[Gilmore(1984)]{14} Gilmore,G 1984 MNRAS 207, 223 
    \bibitem[Bahcall(2000)]{15} Bahcall, N.~A. 2000  Allen's Astrophysical Quantities, 4th ed.,edit. A.N.Cox (Springer:NY) 
    \bibitem[Scott et al (2000)]{16} Scott,D. et al 2000   in   Allen's Astrophysical Quantities, 4th ed.,edit. by A.N.Cox (Springer: NY) 
    \bibitem[Ebert(1955)]{17} Ebert, R. 1955 Z. Astrop. 37,217
   \bibitem[Bonnor(1956)]{18}Bonnor, W.~B. 1956 MNRAS 116,351
   \bibitem[Corbelli \& Palucci(2000)]{CS} Gorbelli, E. \& Salucci, P. 2000 MNRAS 311, 411
   \bibitem[Kam, et al (2015)]{KCC} Kam, Z.S., Carignan, C., Chermin, L., Amnam, P., and Epinat, B. 2015 MNRAS 449, 4048
    \bibitem[Chandrasekhar(1950)]{19} Chandrasekhar, S. 1950  Radiative Transfer (Clarendon: Oxford)
     \bibitem[Rybicki \& Lightman(1979)]{20} Rybicki G.~B.  \&  Lightman,A.~P. 1979  Radiative Processes In Astrophysics (Wiley: New York)
      \bibitem[Mandelbaum et al(2006)]{21} Mandelbaum, R. et al 2006 MNRAS 368, 715 
    \bibitem[Helfer(2017)]{22} Helfer, H.~L. 2017 arXiv: gen-ph/1705.08746 
    \bibitem[Rubin(1980)]{K1} Rubin, V.~C. 1980 \apj 238, 471   
    \bibitem[Rubin(1983)]{K2}Rubin,V.~C. 1983 Science 220, 1339
    \bibitem[Fich \& Tremaine(1991)]{K3} Fich,M. \& Tremaine, S. 1991 ARAA 29,409
    \bibitem[ de Zeeuw \& Franz(1991)]{K4} de Zeeuw,T. \& Franz, M. 1991 ARAA 29, 239
    \bibitem[Kulessa \& Lynden-Bell(1992)]{K5} Kulessa,A.~S. \& Lynden-Bell, D. 1992 MNRAS 255, 105
    \bibitem[Sofue \& Rubin(2001)]{K6} Sofue, A. \& Rubin, V.~C. 2001 ARAA 39, 137
    
    %
 
 \end{thebibliography}
     \end{document}